# Compressible Flow at High Pressure with Linear Equation of State

**William A. Sirignano†**

Department of Mechanical and Aerospace Engineering, University of California, Irvine, CA 92697, USA

Compressible flow varies from ideal-gas behavior at high pressures where molecular interactions become important. Density is described through a cubic equation of state while enthalpy and sound speed are functions of both temperature and pressure, based on two parameters, $A$ and $B$, related to intermolecular attraction and repulsion, respectively. Assuming small variations from ideal-gas behavior, a closed-form solution is obtained that is valid over a wide range of conditions. An expansion in these molecular-interaction parameters simplifies relations for flow variables, elucidating the role of molecular repulsion and attraction in variations from ideal-gas behavior. Real-gas modifications in density, enthalpy, and sound speed for a given pressure and temperature lead to variations in many basic compressible flow configurations. Sometimes, the variations can be substantial in quantitative or qualitative terms. The new approach is applied to choked-nozzle flow, isentropic flow, nonlinear-wave propagation, and flow across a shock wave, all for the real gas. Modifications are obtained for allowable mass-flow through a choked nozzle, nozzle thrust, sonic wave speed, Riemann invariants, Prandtl's shock relation, and the Rankine-Hugoniot relations. Forced acoustic oscillations can show substantial augmentation of pressure amplitudes when real-gas effects are taken into account. Shocks at higher temperatures and pressures can have larger pressure jumps with real-gas effects. Weak shocks decay to zero strength at sonic speed. The proposed framework can rely on any cubic equation of state and be applied to multicomponent flows or to more-complex flow configurations.

† Email address for correspondence: sirignan@uci.edu

## 1. Introduction

The goal of this work is to analyze the differences at high pressures between real-gas compressible-flow behavior and ideal-gas compressible-flow behavior. Specifically, the focus is on canonical, "textbook" theories for compressible flow and the modifications of the classical relations to account for real-gas behavior: one-dimensional, isentropic flow through a choked nozzle; the Riemann invariants for wave propagation; the Prandtl shock relation; and Rankine-Hugoniot relation. As an important feature of the analysis, a linearization of the cubic equation of state (EoS) in parameter space provides a simplifying approximation that facilitates analysis and computation of real-gas flows. This linearization does maintain nonlinear relations amongst the various flow variables and the associated key physics.

Interest in gaseous flows at pressures several-fold above critical pressures is increasing. Decades ago, experimental and computational analysis of flow through choked nozzles was motivated by development of hypersonic wind tunnels. Examples are the studies by Tsien (1946), Donaldson & Jones (1951), and Johnson (1964). More recently, propulsion and power systems are driven towards substantially higher pressures to gain efficiency. Rocket combustors are operating at pressures at hundreds of bars, with the gas generator for propellant turbopumps at even higher pressures. Gas-turbine-engine design is trending towards to peak pressures around sixty bars and diesel engines have long operated at these high peak pressures. Airbag operation involves rocket-level pressures in a small combustion chamber. Of course, other applications related to blasts and industrial processing can exist. In the pioneering works on choked nozzles, the equations of state (EoSs) used at that time are now out-of-date; improved models, although still descendants of the Van der Waal's cubic EoS, now exist. (Chueh & Prausnitz 1967*a*,*b*; Soave 1972)

### 1.1. *Consequence of Real Gas Behavior on Compressible Flow*

The potential for important quantitative differences for inviscid compressible flows between ideal-gas flows and real-gas flows has been well established in the literature. There have been earlier attempts to determine the jump in flow variables across a shock wave. Tao (1955) calculated jumps across normal shocks in Freon-12 flow. The results show significant variations from ideal-gas behavior for shocks with high pressure ratios.

For a pressure ratio equal to 25, the downstream density was about 15% higher for the real gas compared to the ideal gas while the real-gas downstream temperature was 25% lower. Shock flows of nitrogen were considered (Wilson & Regan 1965) where the upstream pressure and temperature varied up to 1000 atmospheres and 2000 K. Correction factors as high as 1.6 for downstream pressure and 1.17 for downstream were found to apply as multiples of the ideal-gas values. The analysis was based on the assumption that the upstream values satisfy the ideal gas law. For the wide range of upstream values considered, this assumption is not acceptable.

For isentropic expansion and compression flows, Tao (1955) plots flow variables versus the Crocco number (Crocco 1958), a nondimensional velocity normalized by the square root of twice the stagnation enthalpy. For the Crocco number in the range of 0.2 to 0.5, they find higher real-gas values compared to ideal-gas values: i.e, 20% for pressure, 10% for density, and 5% for temperature.

For a convergent-divergent nozzle with a standing shock in the divergent (supersonic) portion, both Arina (2004) and Jassim & Muzychka (2008) show significant (i.e., 10 % or more) differences in flow properties for the ideal gas and the real gas. The shock location is also modified. Similar magnitudes of differences are shown by Arina (2004) for the shock tube problem with travelling shock, expansion wave, and contact surface.

Donaldson and Jones performed experiments to measure the ratio of pressure at the choked throat of a nozzle to the stagnation pressure for air flow. They also measured the speed of sound in nitrogen at high pressures and made comparisons using the Beattie-Bridgeman EoS and van der Waal EoS (Poling *et al.* 2001). Johnson used the Beattie-Bridgman EoS to calculate mass-flow rates through a choked nozzle at high stagnation pressures for seven different gases. He found a few percent difference between ideal-gas mass flux and real-gas mass flux, e.g., about a 3.5 % defect for the real-gas nitrogen at $550^o$ R and 100 bar. Ascough (1968) calculated nozzle flow using tabulated thermodynamic data at supply pressures up to 10 bar and temperatures in the 270-400 K range. His results varied from ideal-gas results no higher than the third significant digit, indicating that, if more interesting results exist, they should be sought outside of this temperature-pressure range. More recently, Kim *et al.* (2008) used multidimensional Reynolds-averaged Navier-Stokes equation to treat flow of hydrogen through a choked nozzle. It was difficult to distinguish between real-gas effects and boundary-layer effects in explaining the reduction of mass flow, especially at the higher Reynolds number.

In some configurations and conditions, corrections due to real-gas effects might involve only an adjustment of a value by a few per cent. However, there are situations where such adjustments can have an extraordinarily large impact. A few percent change in thrust resulting from flow through a choked nozzle can have important integrated consequence, for example, on a vehicle-trajectory prediction. As another example, rocket solid propellant or automobile airbag solid explosive typically burns according to a law that gaseous mass generation rate $\dot{m}$ follows pressure $p$ to the power of $n$ with outflow from a pressurized chamber through a choked throat; i.e., $\dot{m} \sim p^n$. If the non-dimensional exponent has the value $n$=0.7 (the top of the practical range) and the discharge coefficient were actually reduced by three-to-ten percent from the design based on an ideal-gas characterization, the chamber pressure would exceed design value by ten-to-thirty-seven percent, creating potentially a very dangerous situation. In addition to this type of case where small corrections have large indirect impact, situations are shown later where a change in a variable due to real-gas correction is large.

### 1.2. *Special Challenges*

The real gas introduces new challenges to the computation of inviscid compressible flows. As noted by Drikakis & Tsangaris (1993), the pressure is no longer primarily a function of the pressure. Rather, it becomes more strongly both a function of pressure and temperature. Enthalpy (or internal energy) becomes related to pressure which creates a new coupling between the energy and momentum equations. Real-gas compressible flow calculations have typically required iteration for a thermodynamic variable involving at least one of the conservation equations. See, for example, the study by Kouremonos (1986) where the energy conservation equation for a jump across a normal shock is used in the iterative process. The real-gas equation of state is typically a cubic algebraic equation with three solutions, two of which can be complex conjugates. Solving the cubic equation, choosing the physically interesting solution, and avoiding the complex numbers form a substantial challenge in the context of intricate flow computations which already demand iterations. Arina (2004) solves three different flow configurations with four different EoSs, the ideal gas EoS and three different real-gas EoSs. The CPU time was always substantially longer for the real gas EoSs. For air flow through a converging-diverging nozzle with a standing shock, the CPU time for the Redlich-Kwong EoS was more than double the perfect gas time; for the same problem with $CO_2$, it was 80% greater. For the shock-tube problem with $CO_2$ where an unsteady expansion and a shock wave occur, the reported CPU time ratio is 3.62. The simpler but less accurate van der Waals EoS is generally less computationally expensive but also less accurate than the Redlich-Kwong EoS. Other real-gas EoSs are more computationally expensive than the Redlich-Kwong EoS. The use of a closed-form approximation to the equation of state can be an extremely good strategy to simplify the complex calculations. See, for example, the comments of Colella & Glaz (1985) on the need for reliable approximations in treating real-gas EoSs.

### 1.3. *Focus and Approach*

Many different types of variations from ideal-gas flow behavior are described as real-gas phenomena. Included are viscous flows, flows with heat and /or mass transport, and flows with various types of relaxation processes such as molecular vibrational excitations, dissociations, a wide variety of other chemical reactions, electronic excitations, and ionization. In this paper, those non-equilibrium processes are not addressed. Here, the focus is on inviscid, compressible flows with equilibrium conditions that do not satisfy the ideal-gas law, $p = \rho RT$ and with enthalpy and internal energy dependent on pressure $p$ as well as temperature $T$. For continuous flows, no non-equilibrium conditions are considered; while for the normal shock wave, the thin zone of $O(10^{-7})$ nanometers in thickness with molecular translational and rotational non-equilibrium is treated as a mathematical discontinuity.

Here, the descriptions *ideal gas* and *perfect gas* are considered to be identical for a gas that satisfies $p = \rho RT$ and undergoes a duration of molecular collision that is negligibly short compared to the mean travel time between consecutive collisions. (This equivalent usage is common in the fields of fluid mechanics and gas kinetic theory but is not accepted in some other fields.) Furthermore, the ideal gas is considered to be calorically perfect (i.e., with constant specific heats). The simplification in the connectivity of the ideal-gas

EoS to the fluid equations for conservation of mass, momentum, and energy can only be fully appreciated after connecting the real-gas EoS to those equations of fluid motion. Treatment of the ideal gas is immensely simplified by the fact that four key quantities are directly proportional to each other: temperature, pressure-to-density ratio, specific enthalpy, and sound speed $c$ squared. Namely, $c^2 = (\gamma - 1)h = \gamma p/\rho = \gamma RT$. Three independent ratios for these four quantities are constant in time and uniform in space for the ideal gas. This should be seen as extremely fortuitous. On the contrary, for a real gas, these ratios can vary significantly. Consequently, findings for an ideal gas which have come to be treated as "law" are known from real-gas analysis to not hold generally.

In the analysis here, a widely accepted form of the cubic EoS is used. Isentropic expansions and compressions are considered; application examples include flow through choked nozzles but nonlinear acoustical wave propagation and modified Riemann invariants are also addressed. The non-isentropic jumps across a shockwave provides another example. A wide range of stagnation temperatures are considered which does have interesting and relevant consequences. An expansion in parameter space is used that more clearly identifies the effects of real-gas molecular interactions, both repulsions and attractions. Discussion is avoided for pressure and temperature domains where two phases or a compressible liquid exist. They deserve special separate attention.

Section 2 describes the basic thermodynamic foundations for the EoS, the enthalpy departure function, and the determination of sound speed. The new linear expansion in the thermodynamic parameters is explained. Using that mathematical linearization, isentropic flow expansions and compressions are analyzed in Section 3. Applications to choked nozzle flow, wave propagation, and shock waves are given in Sections 4, 5, and 6, respectively. Concluding remarks follow in Section 7.

## 2. The Thermodynamic Foundation for Compressible Flow at High Pressures

As a basis for analysis of compressible flow, three quantities which appear in the equations of motion directly or implicitly must be related: enthalpy $h$, pressure-to-density ratio $p/\rho$, and sound speed $c$. The following subsections discuss the more-complex real-gas behavior because of the importance of intermolecular forces at high pressures and densities. Still, in the analysis here, certain effects such as molecular vibration and dissociation are neglected.

### 2.1. *Real-gas Equation of State*

For analysis of compressible flow at very high pressures, corrections to the ideal-gas relations are needed in the supporting thermodynamics theory. Amongst other issues, many classical relations no longer apply in their original forms. In particular, adjustments are needed for the equations of state that describe density (or specific volume), enthalpy, and sound speed as functions of pressure, temperature, and composition.

Poling *et al.* (2001) presents several equation-of-state formulations, including the well known cubic EoSs by Van der Waals, Peng and Robinson, and Redlich and Kwong, governing the compressibility factor $Z \equiv pv/(R_{\mathrm{u}}T) = p/(\rho RT)$. Variations of Redlich-Kwong EoS have been advanced by Chueh & Prausnitz (1967*a*,*b*) and by Soave (1972). For the ideal gas, $Z = 1$ everywhere while, for a real gas, it may vary with space and time. Like any EoS in a system together with conservation equations, the cubic equation presents molar specific volume $v$ or mass density $\rho$ as an implicit function of pressure $p$ and temperature $T$. In addition, an enthalpy departure function gives the difference between the enthalpy for the ideal gas and the enthalpy for the real gas at any given pressure and temperature. Essentially, there is a pair of state equations, one for density and another for enthalpy $h$. The developments proposed here may be done with any of these cubic EoSs but only one is chosen here. In particular, the analysis proceeds with the Soave-Redlich-Kwong (SRK) cubic equation of state which is known for accuracy over a wide range of important applications. For example, Lapuerta & Agudelo (2006) compared several real-gas EoSs with experimental results relevant to Diesel-engine combustion and showed that the SRK EoS gave the best agreement.

The SRK EoS for a single-component fluid is

$$Z^3 - Z^2 + (A - B - B^2)Z - AB = 0 \tag{2.1}$$

where

$$Z \equiv \frac{pv}{R_\mathrm{u}T} = \frac{p}{\rho RT} \tag{2.2}$$

$$A \equiv \frac{ap}{(R_\mathrm{u}T)^2} \tag{2.3}$$

$$B \equiv \frac{bp}{R_\mathrm{u}T} \tag{2.4}$$

$$a \equiv 0.42748\frac{(R_\mathrm{u}T_\mathrm{c})^2}{p_\mathrm{c}}[1 + S(1 - T_\mathrm{r}^{0.5})]^2 \tag{2.5}$$

$$b \equiv 0.08664\frac{R_\mathrm{u}T_\mathrm{c}}{p_\mathrm{c}} \tag{2.6}$$

$$T_\mathrm{r} \equiv \frac{T}{T_\mathrm{c}} \tag{2.7}$$

$$S \equiv 0.48508 + 1.5517\omega - 0.15613\omega^2 \tag{2.8}$$

$R$ and $R_\mathrm{u}$ are the specific and universal gas constants, respectively. Subscript c denotes a thermodynamic critical value . The coefficients $a$ and $b$ (and therefore $A$ and $B$) relate respectively to intermolecular attraction and repulsion. The second and third constant coefficients in the polynomial for $S$ differ slightly from the original Soave (1972) values. They are updated values by Graboski & Daubert (1978) which were also used by Meng & Yang (2003). (A different functional form is recommended for hydrogen; while discussion of gases with differing mathematical description is omitted to avoid distraction from the main themes, the approach is easily extendable to consider them.) In the domain of pressure and temperature where both gas and liquid exist in equilibrium, there is a solution of Equation (2.1) for each phase; thus, two different, physically meaningful $Z$ values can result. Since $p$ and $T$ are identical for each phase, the implication is that there are two values for the specific molar volumes $v_\mathrm{g}$ and $v_\mathrm{l}$ and thereby for the mass densities $\rho_\mathrm{g} = W/v_\mathrm{g}$ and $\rho_\mathrm{l} = W/v_\mathrm{l}$. $W$ is the molar mass. A range of values is considered for $p$ and $T$ where only one phase exists and therefore only one interesting solution to the cubic equation exists. (Complex roots are ignored.) At supercritical conditions, there are ranges of $p$ and $T$ where a compressible fluid exists without discontinuities in properties and may still be labeled as a gas. Thus, reference to $\rho$ and other properties are made with the understanding they apply to a gas.

The properties for flows of gaseous mixtures are not calculated here; however, the extension for that situation is straightforward (Poling *et al.* 2001).

Table 1 presents critical temperature $T_c$, critical pressure $p_c$, acentric factor $\omega$, ratio of specific heats for the ideal gas $\gamma$, and molecular mass $W$ for selected gases. Monatomic, diatomic, and triatomic species are considered. In the calculations in the following sections, argon, nitrogen, and carbon dioxide are analyzed. $\gamma, c_p$, and $c_v$ are values pertaining only to the ideal-gas EoS. For example, as shown by the Equation (B-1), $c_p$ will not be the partial derivative of enthalpy with respect to temperature for the real gas. It will be that derivative only for the ideal gas and it retains only that meaning when used in the real-gas enthalpy relation.

Table 1: **Values for critical temperature, critical pressure, acentric factor, and ratio of specific heats.**

| Gas | $T_c$ (K) | $p_c$ (kPa) | $\omega$ | $\gamma$ | $W$ |
|---|---|---|---|---|---|
| Argon | 150.8 | 4780 | 0 | 1.667 | 40.0 |
| Nitrogen | 126.2 | 3390 | 0.040 | 1.400 | 28.0 |
| Oxygen | 154.6 | 5050 | 0.022 | 1.400 | 32.0 |
| Carbon Dioxide | 304.25 | 7380 | 0.228 | 1.286 | 44.0 |
| Water Vapor | 647.1 | 22064 | 0.344 | 1.333 | 18.0 |

There is no obvious way to reduce the mathematical descriptions of different gases to a similar form for ease of calculation. For example, even if pressure and temperature are normalized and $p/p_c$ and $T/T_c$ are treated, the gases differ significantly through three other parameters in the table.

## 2.2. *Linearized Treatment of Real-gas Equation of State*

The cubic EoS can be solved exactly for the compressibility factor $Z$ in terms of the parameters $A$ and $B$. Five hundred years ago, mathematicians S. del Ferro and N.

Tartaglia obtained solutions for the cubic equation in a form involving cubic roots of functions formed as an algebraically elaborate collection of coefficients in the original equation. The exact solution is not useful in flow analysis. Firstly, the dependency of the coefficients of the cubic equation on unknown variables such as pressure, temperature, and (for multicomponent flows) composition creates a higher-order system; i.e., there are couplings with other differential equations. Thereby, an iterative approach is required. Secondly, in a flow which develops in space and / or time, it is preferred to iterate about the solution at a prior time-step or mesh point rather than returning to decide which of the cubic-equation solutions to choose. The known additional computational challenges for real-gas, compressible flow computation and the need for reliable approximations were addressed in 1.2 with references to Arina (2004), Colella & Glaz (1985), Drikakis & Tsangaris (1993), and Kouremonos (1986).

Approximations through perturbation expansions have been attempted. Tsien (1946) used a linearization concept which differs from this work in several key features. The modern cubic equations were not available at that time; he used the van der Waal's EoS where the parameter $a$ was a constant whereas the modern versions have $a(T)$, i.e., a function of temperature $T$, as shown by Equation (2.8). Also, the important effects of internal energy and enthalpy departures from the ideal gas behavior were missing; among other things, there was no dependence of enthalpy or internal energy upon pressure. Tao (1955) used the Beattie-Bridgeman EoS and perturbation theory with six small parameters to describe the thermodynamics for normal shock analysis. Anand (2012) reports solutions for real-gas shock waves propagating at subsonic velocity. He uses an EOS with a repulsive molecular parameter but without an attractive molecular parameter. Glaister (1988) and Guardone & Vigevano (2002) discuss linearization of the numerical algorithms in Riemann problem solvers.

Here, it is shown that for a wide range of practical interest where the fluid temperature is well above the critical value, certain approximate solutions to the cubic equation of state give sufficient accuracy and clarity of the physics. Often the parameters $A$ and $B$ have magnitudes substantially smaller than unity, making $Z-1$ also small in magnitude. As one example, consider an application by Jorda-Juanos & Sirignano (2016) where, for a mixture typical in combustion of methane, the temperature exceeds 400 K, the parameters $z \equiv Z - 1, A$, and $B$ are small in magnitude. Specifically, they remain $O(10^{-1})$ or less at pressures up to 100 bar. Thereby, a linear perturbation expansion in the three parameters $z, A, B$ provides useful, accurate, simplified algorithms. In particular, neglecting squares and cubes of $z$ and products and squares of $A$ and $B$, Equation (2.1) can be simplified.

$$Z \approx 1 + B - A \tag{2.9}$$

$a$ and therefore $A$ represent the effects of intermolecular attraction; so, an increase in $A$ by itself would increase density at fixed temperature, pressure, and composition. On the other hand, $b$ and therefore $B$ represent the effects of intermolecular repulsion; so, an increase in $B$ by itself at otherwise fixed conditions would decrease density. The linear form of the EoS given by Equation (2.9) explains the impact of the molecular physics on the continuum properties in much clearer fashion than the original cubic EoS. Of course, in addition, it simplifies the flow analysis and carries the dependence on the molecular parameters in a much more informative fashion. In Appendix A, it is shown that the accuracy of that approximation in Equation (2.9) can be maintained over a substantial parameter domain of interest. While density depends simply on $B-A$ here, the enthalpy and sound speed depend on $A$ and $B$ in different ways as well as depending on derivatives of $a(T)$ introduced through the functions $A'$ and $A''$ which are defined in Appendix B where the real-gas enthalpy relation is presented.

If the linear perturbations in Equation (2.9) are neglected, the ideal gas equation, i.e., $Z = 1$, is obtained. This linear form does not apply for two-phase domains or for the supercritical domain where a compressible-liquid behavior occurs. The above-mentioned lower bound on temperature avoids these regions. At some temperatures and pressures, $Z-1$ might be small but $A$ and $B$ might not be sufficiently small to neglect higher-order terms in the cubic equation; thus, bounds on $A$ and $B$ individually are necessary here.

While other forms of the cubic EoS, including the original van der Waals equation and the more recent Peng-Robinson form of the cubic equation, differ from the Redlich-Kwong form presented here, they produce exactly the same linear approximation as given by Equation (2.9). There is only a difference with the Peng-Robinson linearized version in the $S$ parameter which affects the value of $A$ but not its order of magnitude. The van der Waal's equation does not contain the added temperature dependence in the "$a$" parameter. Consequently, the definition of $A$ in the linear version is different. In summary, our analysis here provides a template for linearization of the Peng-Robinson model and several other similar models.

The linear solution to the cubic equation reduces computational effort in flow-field calculations. In the context of the flow calculation, exact solution to the cubic equation requires an iterative process. The solution to a system of differential equations already requires iteration of some kind to get a well converged solution as time is advanced at each step. If at each point in time and space another iteration were added within the finite-difference (or finite-volume) iteration, a substantial increase in computational resources is needed.

Further analysis and development of the linearization is provided in the appendices. In Appendix A, a comparison is made between the exact solution and the linear solution of the cubic equation of state. It is shown there that the linear solution can accurately give the solution to the cubic equation which yields the lowest density and therefore is most applicable to compressible flow. The parameter domain where the accuracy is acceptable is identified. It is also shown in Appendix A that an iterative approach can yield a higher-order solution which has greater accuracy and yields a much larger domain of validity. The linear treatment is extended to the enthalpy and the speed of sound in Appendix B and Appendix C, respectively.

## 3. Isentropic Expansions and Compressions

In this section, the various thermodynamic variables and the velocity are determined as functions of pressure. Isentropic relations are used; thus, stagnation values are fixed. Results are presented in a form that uses stagnation pressure and temperature for normalization. Thus, for an ideal gas, the non-dimensional results do not vary with these stagnation quantities; however, for a real gas, variations occur. These relations describe isentropic expansions and compressions. In our analysis, the entropy value is implicitly determined by the choices of stagnation temperature $\hat{T}$ and stagnation pressure $\hat{p}$ values. Then, it remains constant throughout the expanding or compressing flow.

Foundations are laid in Appendix D where the linear relations are used to relate specific functions to pressure for isentropic and isoenergetic variations in a flow. Specifically, density, enthalpy, temperature, velocity, and sound speed are determined as functions of pressure. While these functions are linear in parameters $A, B$, and their derivatives, the dependencies on pressure are strongly nonlinear.

### 3.1. *Results for isentropic flow*

Figure 1 shows argon flow, nitrogen flow, and carbon dioxide flow results with non-dimensional values for $\rho$ and velocity $u$ as functions of the normalized $p$ for six cases, two for each gas. Comparisons are made with ideal-gas results. Although all calculations here involve temperatures and pressures in a range that extends from above to well above standard conditions including values above the critical values, the cases of sub-figures 1a,c,e are identified as "lower temperature" cases while sub-figure 1b,d,f present the "higher temperature" cases. For the higher temperature cases, the maximum $Z$ occurs at the highest pressure of 30 MPa. The normalized density is higher for the real gas than for the ideal gas; however, the two stagnation densities used for normalization differ by the factor $\hat{Z}$, which is the compressibility factor at the stagnation condition. (Pressure, temperature, enthalpy, velocity, and sound speed are all normalized by the same values for the real and ideal gases; normalization of density is different.) Thus, the dimensional real-gas density is actually lower than the ideal-gas value. For example, for nitrogen in sub-figure 1d, the normalized density is a few percent higher for the real gas. However, $\hat{Z} = 1.097$; thus, the real-gas dimensional density is about eight percent lower than the ideal-gas value in the mid-range of the expansion. For the lower temperature cases, the value of $\hat{Z}$ can fall below the ideal-gas value. For example, with carbon dioxide in sub-figure 1e, the real-gas normalized density is a few percent higher than the ideal-gas normalized density. However, here $\hat{Z} = 0.901$ as shown in Table 2. Thus, the stagnation density for the real-gas is eleven per cent higher than for the ideal gas and the dimensional real-gas density exceeds the ideal-gas density by about fifteen per cent in the mid-pressure range. In this lower temperature range, the attractive molecular forces that influence $A$ are more effective than the repulsive forces that influence $B$. The effect of the variation of $\hat{Z}$ from the ideal-gas value is most significant here on density. At high pressures, the intermolecular forces prevent the increase of density in proportion to the pressure as would occur with the ideal gas. In the next section, the consequence for mass flow through a choked nozzle is shown.

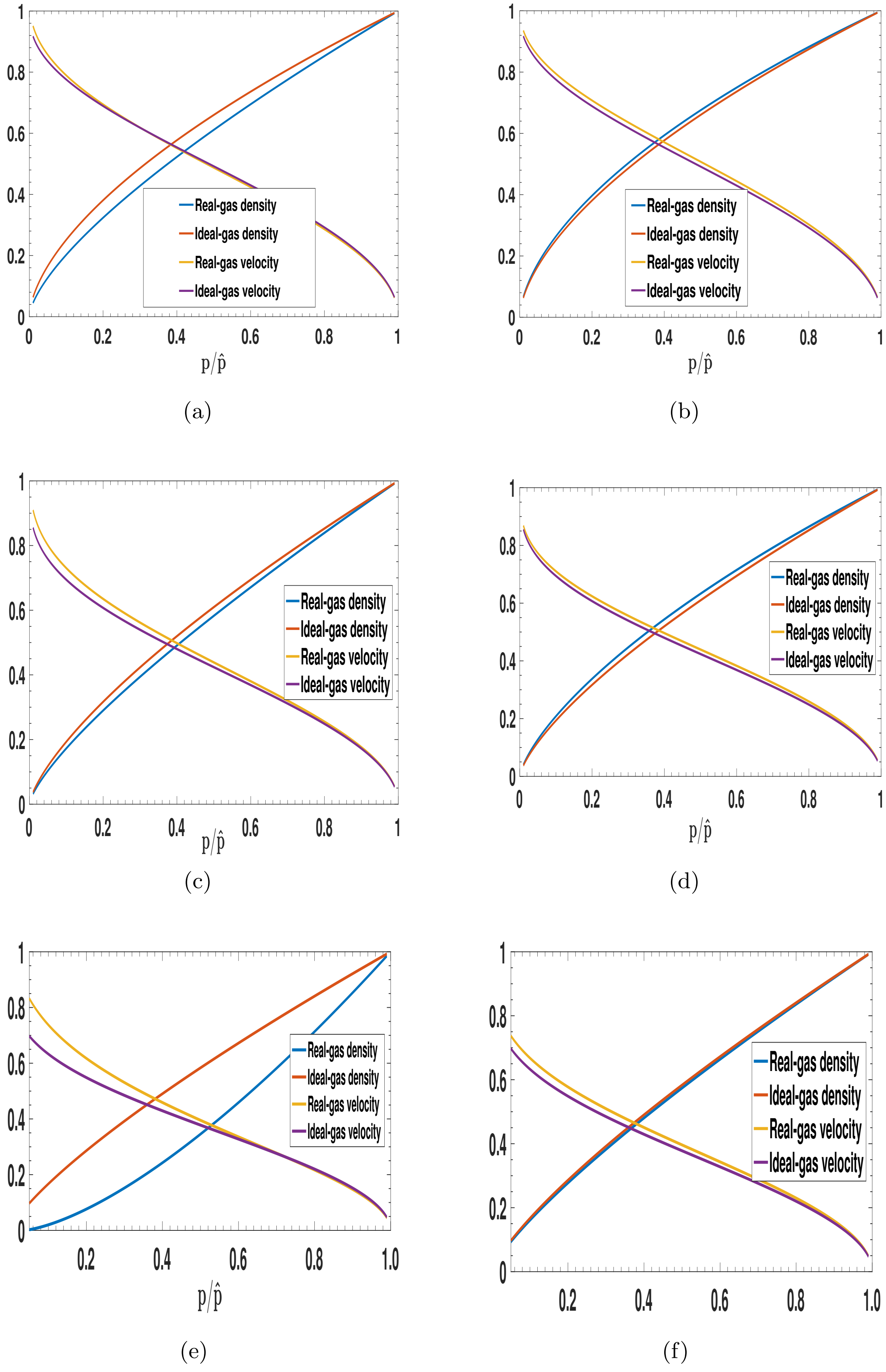

**Fig. 1 Solutions for non-dimensional density and velocity versus non-dimensional pressure. (a) Argon, 300 K, 10 MPa; (b) Argon, 1000 K, 30 MPa; (c) Nitrogen, 400 K, 12 MPa; (d) Nitrogen, 1000 K, 30 MPa; (e) Carbon dioxide, 450 K, 10 MPa; (f) Carbon dioxide, 1000 K, 30 MPa.**

Table 2: **Comparison between ideal-gas flow and real-gas flow: compressibility factors and ratios for stagnation enthalpy and mass-flux.**

| Gas | $\hat{T}$ (K) | $\hat{p}$ (MPa) | $\hat{Z}$ | $Z_t$ | $\dot{m}_{real}/\dot{m}_{ideal}$ | $\hat{h}/(c_p\hat{T})$ |
|---|---|---|---|---|---|---|
| *Argon* | 300 | 10 | 0.946 | 0.907 | 1.051 | 0.902 |
| *Argon* | 1000 | 30 | 1.079 | 1.044 | 0.961 | 1.030 |
| $N_2$ | 400 | 12 | 1.047 | 1.013 | 0.982 | 0.999 |
| $N_2$ | 1000 | 30 | 1.097 | 1.059 | 0.957 | 1.028 |
| $CO_2$ | 450 | 10 | 0.901 | 0.900 | 1.237 | 0.923 |
| $CO_2$ | 1000 | 30 | 1.090 | 1.047 | 0.959 | 1.020 |

The real-gas velocity generally exceeds the ideal-gas value for both the higher and lower stagnation temperature cases. Accordingly, Figure 2 shows that the real-gas enthalpy decreases faster than the ideal-gas enthalpy with decreasing pressure. The stagnation enthalpy for the same stagnation temperature differs between the real and ideal gases.

As Table 2 indicates, the cases at higher stagnation temperature generally have real-gas enthalpy exceeding the ideal-gas enthalpy while the cases at lower stagnation temperature sometimes have real-gas enthalpy values below the ideal-gas enthalpy values. The kinetic energy that can manifest from an expansion increases with increasing stagnation enthalpy.

Enthalpy and temperature for isentropic flows of argon, nitrogen, and carbon dioxide are presented in figure 2 as functions of the non-dimensional pressure for both the real and ideal gases. Generally, the normalized real-gas enthalpy differs from the normalized temperature for the real gas by a few percent for those conditions. The normalized temperatures differ less significantly for the real gas and ideal gas. By construction, normalized temperature equals normalized enthalpy for the ideal gas. At higher stagnation temperatures, the real-gas enthalpy is higher than either the real-gas temperature or the ideal-gas enthalpy (temperature) for the three gases as seen in sub-figures 2b,d,f. This implies that there is more energy in the real gas at those same conditions and explains the larger velocity obtained for the real gas in an isentropic expansion. However, a reversal might occur at lower stagnation temperatures as seen in 2a,c; the ideal-gas enthalpy and temperature values can exceed those for the real gas.

At higher stagnation temperature, the real gas has a higher sound speed than the ideal gas at the same temperature and pressure. At lower stagnation temperatures, the relative magnitudes can be reversed.

## 4. One-dimensional Nozzle Flow

The first example of a compressible flow to be studied is isentropic flow through a choked nozzle. One-dimensional steady flow is examined. The results of the previous section can be applied.

For an ideal gas, steady-state or quasi-steady-state flow has the Mach number $M$ at the nozzle entrance determined by the ratio of the specific heats $\gamma \equiv c_{\mathrm{p}}/c_{\mathrm{v}}$ and the ratio of the nozzle-entrance cross-sectional area to the nozzle-throat cross-sectional area $A_{\mathrm{e}}/A_{\mathrm{t}}$. Detailed analysis of one-dimensional compressible flow is in many references, e.g., Crocco (1958), Liepmann & Roshko (1957), and Saad (1993). The relation is not as simple for a non-ideal gas described by the cubic equation of state (2.1).

Consider here a one-dimensional, steady, inviscid flow without body forces. For a constant mass flux $\dot{m} = \rho u A$, the continuity and momentum relations are

$$\frac{d\rho}{\rho} + \frac{du}{u} + \frac{dA}{A} = 0 \; ; \tag{4.1}$$

$$\rho u du + dp = 0 \tag{4.2}$$

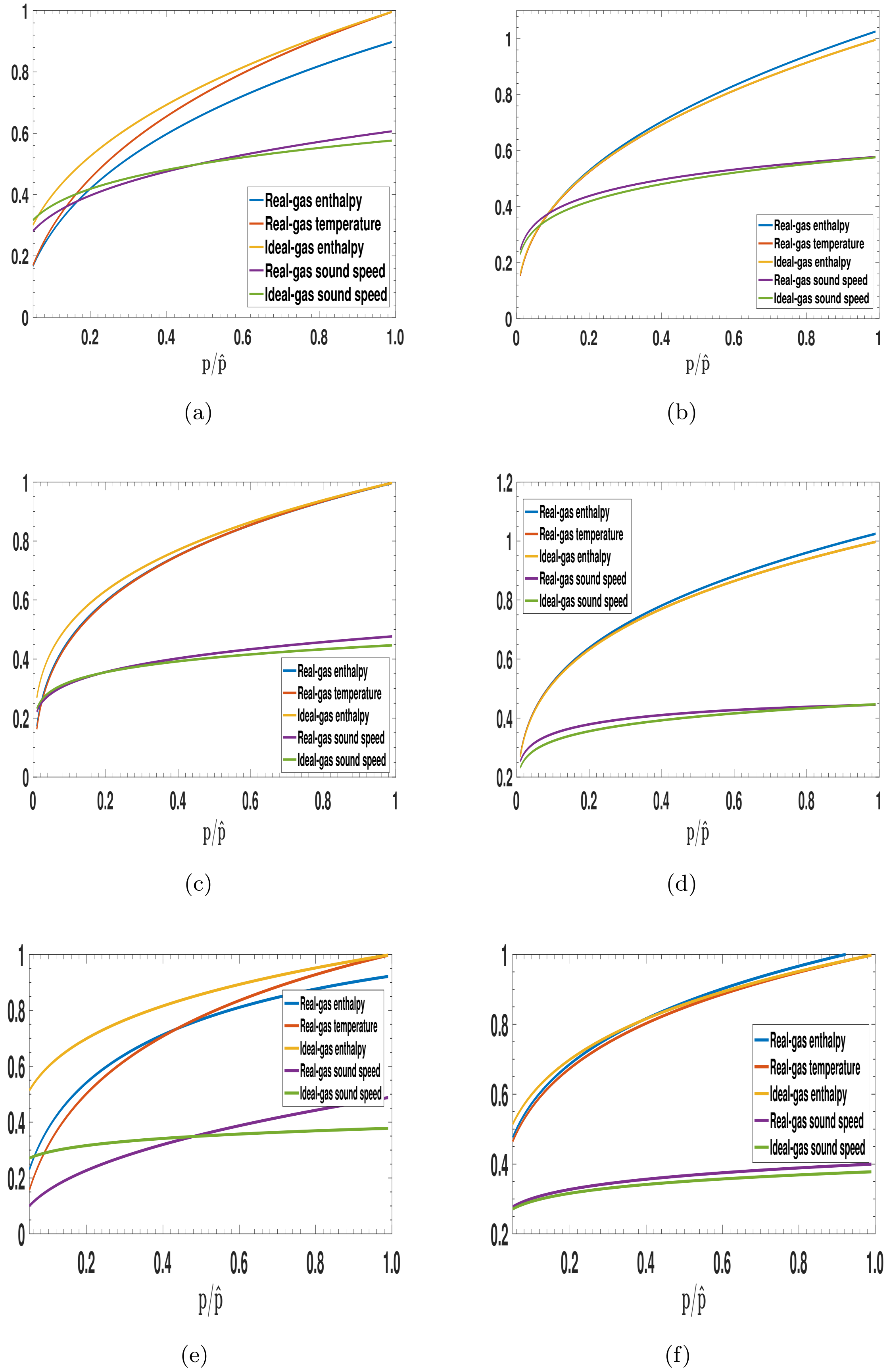

**Fig. 2** **Solutions for non-dimensional enthalpy, temperature, and sound speed versus non-dimensional pressure. (a) Argon, 300 K, 10 MPa; (b) Argon, 1000 K, 30 MPa; (c) Nitrogen, 400 K, 12 MPa; (d) Nitrogen, 1000 K, 30 MPa; (e) Carbon dioxide, 450 K, 10 MPa; (f) Carbon dioxide, 1000 K, 30 MPa.**

From Equations (4.2), it can be shown that

$$\frac{u}{A}\frac{dA}{du} = \frac{u^2 - \frac{dp}{d\rho}}{\frac{dp}{d\rho}} \tag{4.3}$$

At the nozzle throat where $A$ reaches a minimum value, the Mach number $M = 1$ and the local throat velocity is given by $u_{\mathrm{t}} = \sqrt{dp/d\rho|_{\mathrm{t}}}$. This result is not based on the isoenergetic or isentropic assumption. So, even in the non-adiabatic case, it holds. Only friction and body force have been neglected.

If there is no chemical reaction, vibrational relaxation, or heat transfer, the flow is isentropic and $dp = c^2 d\rho$. Then, Equation (4.3) yields

$$\frac{u}{A}\frac{dA}{du} = \frac{u^2 - c^2}{c^2} = M^2 - 1 \tag{4.4}$$

Now, the throat velocity $u_{\mathrm{t}} = \sqrt{\partial p/\partial\rho|_{\mathrm{s=constant,t}}} = c_{\mathrm{t}}$ where $c_{\mathrm{t}}$ is the throat value of the thermodynamic variable $c$ which is the speed of sound, as shown by Equation (C-1). Thereby, for a steady, inviscid, isoenergetic, homocompositional flow, sonic velocity occurs at the throat. This conclusion has not been constrained by any assumption about the equations of state for density and enthalpy. The Bethe-Zel'dovich-Thompson fluid (Kluwick 1993), which is outside our immediate interest, can have sonic flow at other positions besides the throat.

For an ideal-gas isentropic flow, the knowledge of the fluid composition and the prescription of stagnation enthalpy (or stagnation temperature) immediately yields the value of temperature at the throat because enthalpy and sound speed depend only on temperature and are independent of pressure. That is, $h^{\mathrm{o}} = h(T_{\mathrm{t}}) + [c(T_{\mathrm{t}})]^2/2$ is a relation fixing $T_{\mathrm{t}}$. Then, with knowledge of the stagnation pressure and use of the polytropic relation applied at constant entropy, the pressure at the throat $p_{\mathrm{t}}$ is obtained. From knowledge of the pressure and temperature at the throat, all other quantities, i.e., $c, h, u$, are easily determined from thermodynamic relations. Thus, once stagnation values for pressure and temperature are prescribed, all values at the nozzle throat are readily determined by algebraic relations without need to integrate Equations (4.2) numerically. This is not possible in the real-gas case where integration of the equations becomes necessary.

In contrast, for the real gas, $h^{\mathrm{o}} = h(p_{\mathrm{t}}, T_{\mathrm{t}}) + [c(p_{\mathrm{t}}, T_{\mathrm{t}})]^2/2$. Thus, specification of the stagnation enthalpy only gives a relation between $p_{\mathrm{t}}$ and $T_{\mathrm{t}}$. Also, there is no polytropic relation between pressure and temperature. So, in general, numerical integration becomes necessary.

### 4.1. *Values at the sonic location*

At the sonic point, where $u = c$ and $u^2 = c^2$, use of Equations (D-13) and (D-17) yields a condition for the pressure value there. Using subscript t for that point and using coefficient definitions from Equation (D-5), it can be stated that

$$\begin{aligned}
\frac{\gamma+1}{2}\big(\frac{p_\mathrm{t}}{\hat{p}}\big)^{(\gamma-1)/\gamma} &= 1+(\lambda_1-\lambda_\mathrm{b})\left[\frac{\gamma-1}{\gamma}\frac{p_\mathrm{t}}{\hat{p}}-\big(\frac{p_\mathrm{t}}{\hat{p}}\big)^{(\gamma-1)/\gamma}+\frac{1}{\gamma}\right] \\
&-\lambda_2\big[(\gamma-1)\big(\frac{p_\mathrm{t}}{\hat{p}}\big)^{1/\gamma}-\big(\frac{p_\mathrm{t}}{\hat{p}}\big)^{(\gamma-1)/\gamma}+2-\gamma\big] \\
&+\lambda_3\big[\frac{2(\gamma-1)}{\gamma+1}\big(\frac{p_\mathrm{t}}{\hat{p}}\big)^{(1-\gamma)/2\gamma}-\big(\frac{p_\mathrm{t}}{\hat{p}}\big)^{(\gamma+1)/\gamma}+\frac{3-\gamma}{\gamma+1}\big] \\
&+\frac{\gamma-1}{2}\big(\frac{p_\mathrm{t}}{\hat{p}}\big)^{(\gamma-1)/\gamma}\Bigg[\lambda_b-\lambda_1+\lambda_2-\lambda_3-2(\lambda_\mathrm{b}-\lambda_1)\big(\frac{p_\mathrm{t}}{\hat{p}}\big)^{1/\gamma} \\
&-\frac{2\lambda_2}{\gamma}\big(\frac{p_\mathrm{t}}{\hat{p}}\big)^{(2-\gamma)/\gamma}+\frac{(5-\gamma)\lambda_3}{2}\big(\frac{p_\mathrm{t}}{\hat{p}}\big)^{(3-\gamma)/2\gamma}\Bigg] \\
&= 1+(\lambda_\mathrm{b}-\lambda_1)\big[\frac{\gamma+1}{2}\big(\frac{p_\mathrm{t}}{\hat{p}}\big)^{(\gamma-1)/\gamma}-\frac{\gamma^2-1}{\gamma}\frac{p_\mathrm{t}}{\hat{p}}-\frac{1}{\gamma}\big] \\
&+\lambda_2\big[\frac{\gamma+1}{2}\big(\frac{p_\mathrm{t}}{\hat{p}}\big)^{(\gamma-1)/\gamma}-\frac{\gamma^2-1}{\gamma}\big(\frac{p_\mathrm{t}}{\hat{p}}\big)^{1/\gamma}+\gamma-2\big] \\
&+\lambda_3\big[\frac{\gamma-1}{2}\big(\frac{p_\mathrm{t}}{\hat{p}}\big)^{(\gamma-1)/\gamma}+\frac{2(\gamma-1)}{\gamma+1}\big(\frac{p_\mathrm{t}}{\hat{p}}\big)^{(1-\gamma)/2\gamma} \\
&-\big(\frac{3-\gamma}{2}\big)^2\big(\frac{p_\mathrm{t}}{\hat{p}}\big)^{(\gamma+1)/\gamma}+\frac{3-\gamma}{\gamma+1}\big]
\end{aligned} \tag{4.5}$$

To zeroeth order, $p_\mathrm{t}/\hat{p} = [2/(\gamma+1)]^{\gamma/(\gamma-1)}$ which may be substituted into the first-order terms. Define

$$\mathit{\Gamma}_0 \equiv \frac{2}{\gamma+1} \quad ; \quad \mathit{\Gamma}_1 \equiv \big(\frac{2}{\gamma+1}\big)^{\gamma/(\gamma-1)} \quad ; \quad \mathit{\Gamma}_2 \equiv \big(\frac{2}{\gamma+1}\big)^{1/(\gamma-1)} \tag{4.6}$$

Then, it is obtained that

$$\begin{aligned}
\frac{p_\mathrm{t}}{\hat{p}} = \mathit{\Gamma}_1\Bigg[&1+(\lambda_\mathrm{b}-\lambda_1)\big[\frac{\gamma-1}{\gamma}-\frac{(\gamma^2-1)}{\gamma}\mathit{\Gamma}_1\big]+\lambda_2\big[\gamma-1-\frac{\gamma^2-1}{\gamma}\mathit{\Gamma}_2\big] \\
&+\lambda_3\big[\big(\frac{2}{\gamma+1}\big)^{1/2}(\gamma-1)-\big(\frac{3-\gamma}{2}\big)^2\mathit{\Gamma}_1\mathit{\Gamma}_2+\mathit{\Gamma}_0\big]\Bigg]^{\gamma/(\gamma-1)}
\end{aligned} \tag{4.7}$$

Now, substitution from Equation (4.7) into Equations (D-7, D-10, D-14, D-17, D-12) allows determination of other variables at the sonic point as functions of $\gamma, \hat{A}$ and $\hat{B}$. The values of $\rho_\mathrm{t}$ and $c_\mathrm{t}$ are especially useful in determining mass flow and thrust for a choked nozzle configuration. For example,

$$\frac{\rho_{\rm t}}{\hat{\rho}} = \big(\frac{p_{\rm t}}{\hat{p}}\big)^{1/\gamma} + (\lambda_1 - \lambda_{\rm b})\big[\mathit{\Gamma}_2^2 - \mathit{\Gamma}_2\big] + \lambda_2\big[\mathit{\Gamma}_2 - \frac{\mathit{\Gamma}_2^2}{\mathit{\Gamma}_0}\big] + \lambda_3\big[\mathit{\Gamma}_2^{5/2}\mathit{\Gamma}_1^{-1/2} - \mathit{\Gamma}_2\big] \quad (4.8)$$

and

$$\frac{c_{\rm t}}{(2c_{\rm p}\hat{T})^{1/2}} = \left(\frac{\gamma-1}{2}\right)^{1/2}\hat{Z}^{1/2}\big(\frac{p_{\rm t}}{\hat{p}}\big)^{(\gamma-1)/2\gamma}\Bigg[1 - \lambda_{\rm b} + \lambda_1 - \lambda_2 + \lambda_3$$
$$+2(\lambda_{\rm b} - \lambda_1)\mathit{\Gamma}_2 + \frac{2\lambda_2}{\gamma}\frac{\mathit{\Gamma}_2^2}{\mathit{\Gamma}_1} - \frac{(5-\gamma)\lambda_3}{2}\frac{\mathit{\Gamma}_2}{\mathit{\Gamma}_0^{1/2}}\Bigg]^{1/2} \quad (4.9)$$

where the lower-order solution for $p_{\rm t}$ has been substituted into the higher-order terms of Equations (4.8) and (4.9).

### 4.2. *Dependence on Mach number*

From Equations (D-13) and (D-17), the Mach number $M$ can be determined as a function of pressure in the one-dimensional isentropic flow. Namely,

$$M = \frac{u}{c} = \left[\left(\frac{2}{\gamma-1}\right)\big(\frac{p}{\hat{p}}\big)^{(1-\gamma)/\gamma}\frac{\big(1 - (\frac{p}{\hat{p}})^{(\gamma-1)/\gamma} + \mathit{\Lambda}_2(\frac{p}{\hat{p}})\big)}{1 + \mathit{\Lambda}_3(\frac{p}{\hat{p}})}\right]^{1/2} \quad (4.10)$$

where the functions $\mathit{\Lambda}_2$ and $\mathit{\Lambda}_3$ are defined by Equations (D-11) and (D-16) and encapsulate the first-order corrections for the real gas.

To lowest order, there is the ideal-gas result $p/\hat{p} = m^{\gamma/(1-\gamma)}$ where $m \equiv 1 + [(\gamma - 1)/2]M^2$. This may be substituted into the higher-order terms in Equation (4.10) to obtain the approximation for pressure as a function of Mach number.

$$\frac{p}{\hat{p}} = \Big[1 + \frac{\gamma-1}{2}M^2[1 + \mathit{\Lambda}_2(m^{\gamma/(1-\gamma)})] - \mathit{\Lambda}_1(m^{\gamma/(1-\gamma)})\Big]^{\gamma/(1-\gamma)} \quad (4.11)$$

where $\mathit{\Lambda}_1$ is defined by Equation (D-8).

### 4.3. *Dependence on cross-sectional area*

From the one-dimensional continuity relation for choked flow through a nozzle, it follows that

$$\frac{\mathrm{A}}{\mathrm{A_t}} = \frac{(\rho_{\rm t}/(\hat{Z}\hat{\rho}))}{(\rho/\hat{\rho})}\frac{\left(c_{\rm t}/\sqrt{2c_{\rm p}\hat{T}}\right)}{\left(u/\sqrt{2c_{\rm p}\hat{T}}\right)} \quad (4.12)$$

Substitution from Equations (D-7, D-14, D-18, 4.7) into Equation (4.12) yields $\mathrm{A}/\mathrm{A}_\mathrm{t}$ as a function of $p/\hat{p}, \gamma, \hat{A}$ and $\hat{B}$. To lowest order, a relation for the ideal-gas flow is given.

$$\frac{\mathrm{A}}{\mathrm{A}_\mathrm{t}} = \big[1+\frac{\gamma-1}{2}M^2\big]^{1/(\gamma-1)}\big(\frac{\gamma-1}{2}M^2\big)^{-1/2}\big[1+\frac{\gamma-1}{2}M^2\big]^{1/2}\Gamma_1^{1/\gamma}\big(\frac{\gamma-1}{\gamma+1}\big)^{1/2}$$
$$= \frac{1}{M}\Big[\frac{2}{\gamma+1}\big(1+\frac{\gamma-1}{2}M^2\big)\Big]^{(\gamma+1)/(2(\gamma-1))} \tag{4.13}$$

To solve Equation (4.12) for $M$ as a function of $\mathrm{A}/\mathrm{A}_\mathrm{t}$, it is convenient to solve prior Equation (4.13) for a first-order approximation of $M$ as a function of $\mathrm{A}/\mathrm{A}_\mathrm{t}$. It yields two solutions; one is supersonic and the other is subsonic. Then, the solution for $M$ from this lower-order analysis can be substituted into the first-order terms of Equation (4.12).

## 4.4. *Mass flux and thrust*

The mass flux through the choked nozzle $\dot{m}$ depends on stagnation pressure, stagnation temperature, ratio of specific heats, and throat cross-sectional area.

$$\frac{\dot{m}}{\hat{p}\mathrm{A}_\mathrm{t}/(R\hat{T})^{1/2}} = \frac{\rho_\mathrm{t} c_\mathrm{t} \mathrm{A}_\mathrm{t}}{\hat{p}\mathrm{A}_\mathrm{t}/(R\hat{T})^{1/2}} \tag{4.14}$$

where the inputs from Equations (4.7, 4.8, 4.9) are made. For the ideal gas, this reduces to

$$\dot{m}_\mathrm{ideal} = \gamma^{1/2}\big(\frac{2}{\gamma+1}\big)^{(\gamma+1)/(2(\gamma-1))}\frac{\hat{p}\mathrm{A}_\mathrm{t}}{(R\hat{T})^{1/2}} \tag{4.15}$$

The product $\rho_\mathrm{t} u_\mathrm{t} = \rho_\mathrm{t} c_\mathrm{t}$ in Equation (4.14) can be determined in several ways: (*i*) Equations (4.8) and (4.9) can be used; (*ii*) the values of $\rho$ and $u$ can be used at the pressure where $u = c$; or (*iii* ) the magnitude of the maximum value of the product $\rho u$ can be determined over the pressure range. The numerical results are close but differences occur because of second-order errors in the linear method. The third approach has arbitrarily been selected.

The thrust force $F$ can also be determined as a function of stagnation properties, values of the variables at the nozzle throat and exit, and cross-sectional area. If the subscripts e and a respectively denote exit values and ambient values, the standard relation is $F = \dot{m}u_\mathrm{e} + (p_\mathrm{e} - p_\mathrm{a})A_\mathrm{e}$. Thus, the non-dimensional thrust is given as

$$\frac{F}{\hat{p}A_\mathrm{t}} = \frac{\rho_\mathrm{t} c_\mathrm{t} u_\mathrm{e}}{\hat{p}} + \big(\frac{p}{\hat{p}} - \frac{p_\mathrm{a}}{\hat{p}}\big)\frac{A_\mathrm{e}}{A_\mathrm{t}} \tag{4.16}$$

The specific impulse is defined as $I \equiv F/(\dot{m}g)$ with units of seconds. A normalized value can be calculated from Equations (4.14) and (4.16).

$$\frac{Ig}{\sqrt{R\hat{T}}} = \frac{F/(\hat{p}\mathrm{A_t})}{\dot{m}(R\hat{T})^{1/2}/(\hat{p}\mathrm{A_t})} \tag{4.17}$$

4.5. *Results for one-dimensional flow*

Figure 3 shows comparisons of Mach number and normalized cross-sectional area versus normalized pressure for both real and ideal flows. Mach number is generally slightly higher for real gases at the lower stagnation temperatures; at higher temperatures, no general behavior appears. This comparison is at a given pressure not a given cross-sectional area. This point is noteworthy because the areas for the real gas and ideal gas at that pressure can differ. Generally, in the supersonic region at lower stagnation temperatures, for the identical Mach number, the real gas has a higher pressure. At a given pressure and lower stagnation temperatures, the real gas has larger cross-sectional area in the supersonic domain.

Figure 4 gives comparisons between the real gas and ideal gas for mass flux $\dot{m}/\dot{m}_{ideal}$, momentum flux $\dot{m}u/(\dot{m}_{ideal}u_{ideal})$, thrust, compressibility factor $Z$, and thrust ratio $F/F_{ideal}$. Real-gas mass flux is generally lower (higher) than ideal-gas flux for flows where $Z-1$ is positive (negative). This rule generally occurs at higher stagnation temperatures with some deviation at lower stagnation temperatures. Sub-figure 4c shows a transitional case for the value of $Z-1$. This higher temperature reduction in allowable mass flux is highly relevant to combustion at high pressures. The compressibility factor generally increases with increasing pressure and increasing stagnation temperature. Previous works (Johnson 1964; Ascough 1968; Kim *et al.* 2008) reported mass flux through choked nozzles for generally low stagnation pressures. With the exception of a case with water-vapor (i.e., steam) flow (Johnson 1964), they found the real-gas flow had a greater mass flux than the ideal-gas flow. It follows that real-gas flow gives less mass discharge at higher stagnation temperatures and the reversal is related to the change in relative magnitudes of the repulsion and attraction parameters.

The momentum-flux ratio and the thrust ratio exhibit similar trends, always decreasing at lower stagnation temperatures with increasing pressure. At optimal thrust, momentum flux and thrust become equal. For the very high stagnation pressures considered, the expansions have not reached a cross-sectional area that gives the designated ambient pressure ( 1 bar). Thus, the momentum-flux ratio exceeds the thrust ratio.

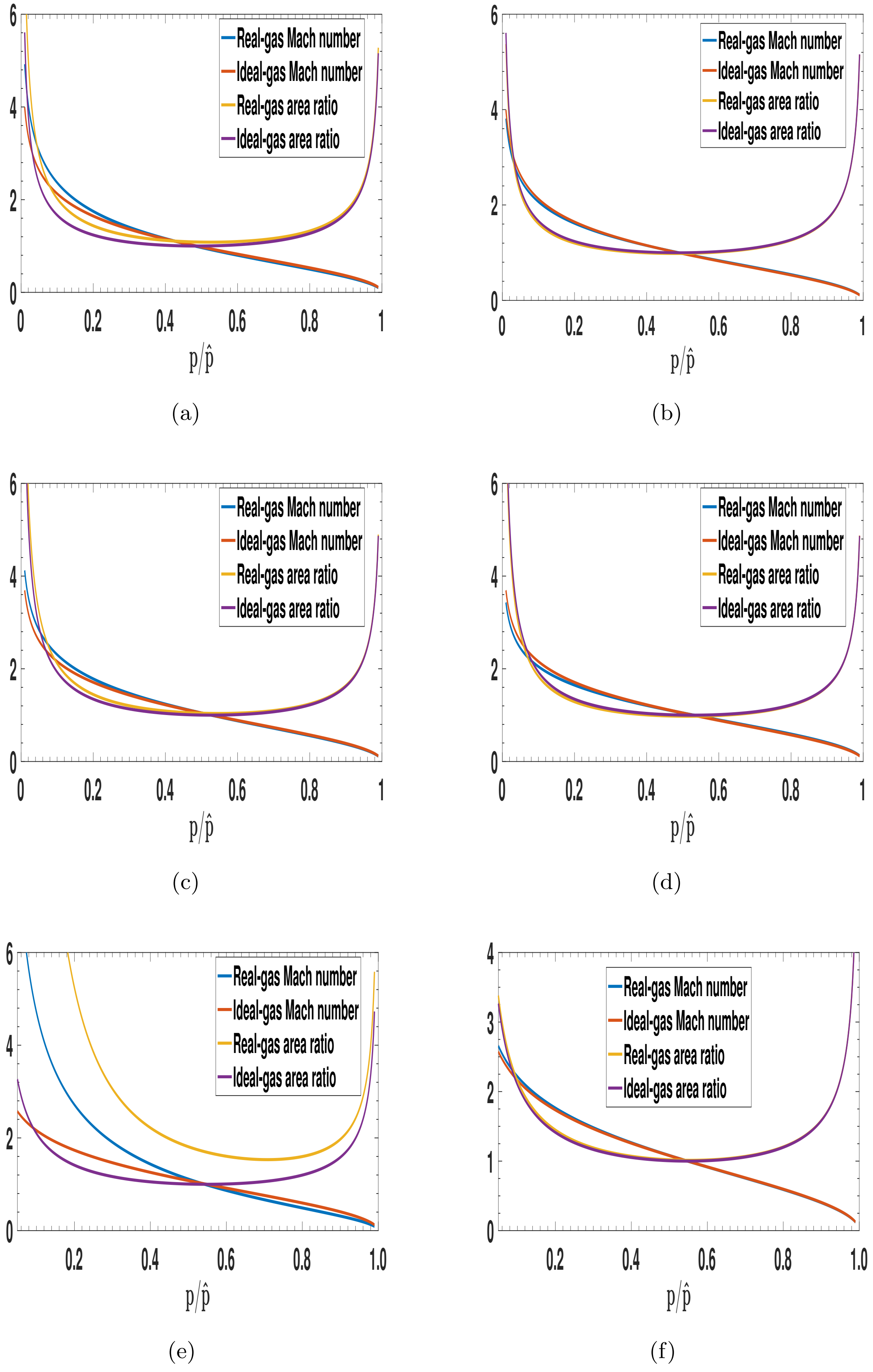

**Fig. 3 Solutions for Mach number and area ratio versus non-dimensional pressure for choked nozzle flow. (a) Argon, 300 K, 10 MPa; (b) Argon, 1000 K, 30 MPa; (c) Nitrogen, 400 K, 12 MPa; (d) Nitrogen, 1000 K, 30 MPa; (e) Carbon dioxide, 450 K, 10 MPa; (f) Carbon dioxide, 1000 K, 30 MPa.**

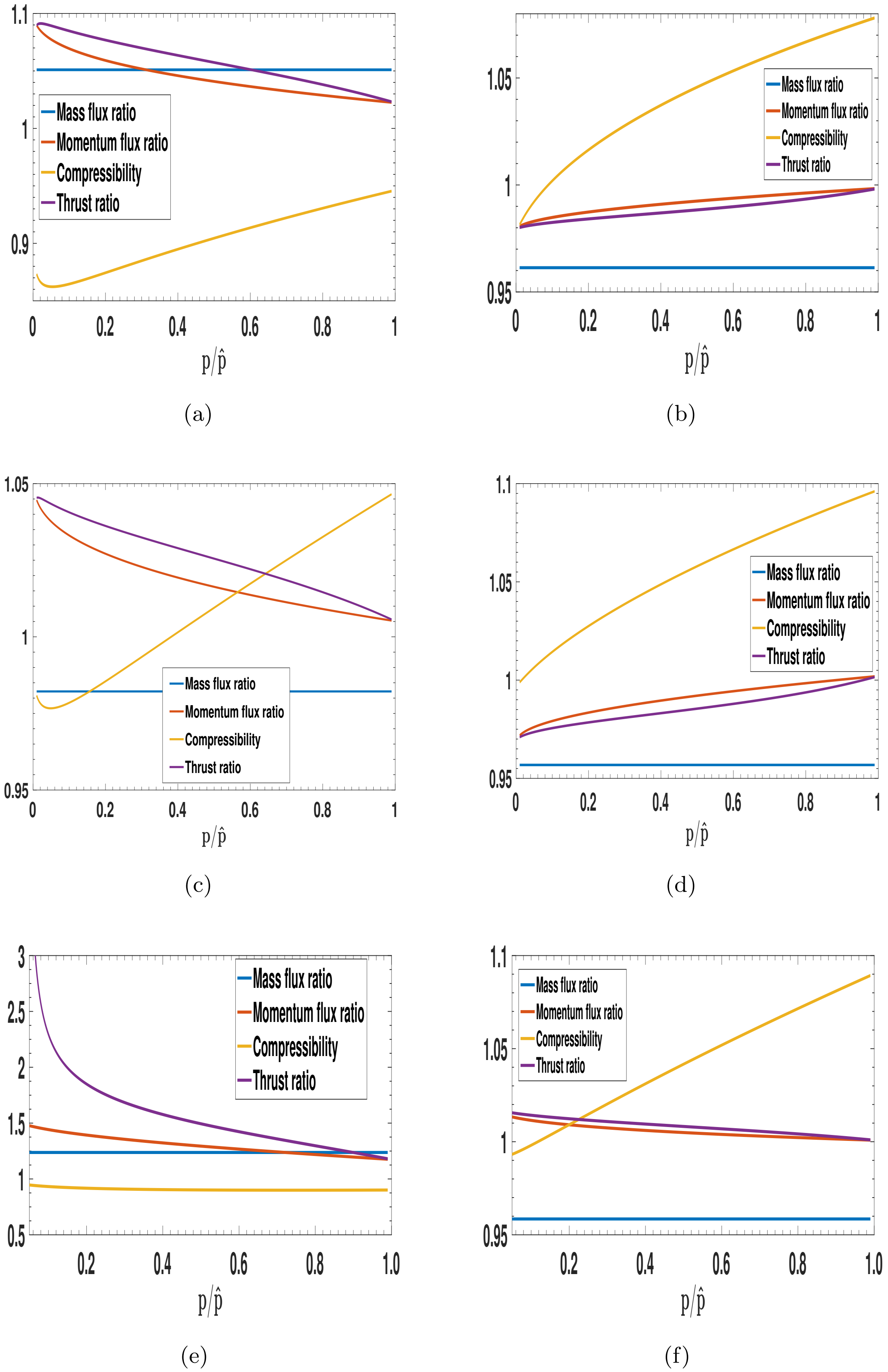

**Fig. 4 Comparison between real-gas nozzle flow and ideal-gas nozzle flow: real-gas-to-ideal-gas ratios of mass flux, momentum flux, and thrust. (a) Argon, 300 K, 10 MPa; (b) Argon, 1000 K, 30 MPa; (c) Nitrogen, 400 K, 12 MPa; (d) Nitrogen, 1000 K, 30 MPa; (e) Carbon dioxide, 450 K, 10 MPa; (f) Carbon dioxide, 1000 K, 30 MPa.**

Table 3: **Optimal thrust values for nitrogen flow through nozzle.**

| $\hat{T}$ (K) | $\hat{p}$ (MPa) | $\dot{m}/\dot{m}_{\text{ideal}}$ | $F/(\hat{p}A_{\text{t}})$ | $F_{\text{ideal}}/(\hat{p}A_{\text{t}})$ |
|---|---|---|---|---|
| 1000 | 30 | 0.957 | 1.557 | 1.604 |
| 1500 | 30 | 0.970 | 1.585 | 1.604 |
| 1500 | 45 | 0.957 | 1.585 | 1.613 |
| 2000 | 45 | 0.971 | 1.630 | 1.613 |

It is interesting to examine the optimal thrust which manifests when the supersonic flow is expanded to ambient pressure. Several combinations of stagnation pressure and temperature were taken with results shown in Table 3 for nitrogen with an expansion to 1 bar. In all cases, the mass flux for the real gas was a few percent below the ideal-gas values. Generally, the real-gas thrust is lower than the ideal-gas thrust at higher stagnation pressures and temperatures.

There is longstanding knowledge that, at high pressures, the allowable mass flux through a choked nozzle can differ by a few per cent between the ideal gas and the real gas. Here, one finds for nitrogen in sub-figure 4c that, at 400 K (720$^o$R) and 120 bar, $Z$= 104.7 and mass-flux defect (i.e., one minus mass-flux ratio) is 0.017. Considering differences in the EoS, these values compare favorably with the values of 103.7 and 0.027 reported by Johnson (1964) for 700$^o$R and 100 atm.

## 5. Wave Motion and Modified Riemann Invariants

If the unsteady flow is homentropic ( i.e., it has uniform and constant entropy value), it is useful to non-dimensionalize the equations using values of pressure and temperature that yield the same value of entropy as the actual flow entropy value. Then, the variation of the dimensional flow variable from the reference value is always isentropic and a relation between the dimensional value and the reference value is readily available. The stagnation values for pressure and temperature at any specified (albeit arbitrary) point in space and time satisfies this need and can serve as the reference values for all space and time. The stagnation pressure and the stagnation temperature are sufficient to determine entropy (given composition). Although the values of that pair (and of other stagnation variables) can individually vary through the flow in space or time, they are coupled for a homentropic flow in such a way as to produce the identical entropy value. Essentially, all paths under consideration here occur at the same entropy value whether they are the actual path of a fluid particle or the abstract path from static to stagnation values.

Consider one-dimensional wave motion with isentropic conditions. Let $p', \rho', u'$, and $c'$ be the normalized quantities given by Equations (D-7, D-14, D-18). Velocity $\sqrt{2c_{\mathrm{p}}\hat{T}}$ and length $L$ are used to provide the normalized $x'$ and $t'$. Here, the chosen reference properties for the non-dimensional scheme are the quiescent conditions for pressure and temperature. The continuity and momentum equation may be written as

$$\frac{\partial(ln\ \rho')}{\partial t'} + u'\frac{\partial(ln\ \rho')}{\partial x'} + \frac{\partial u'}{\partial x'} = 0 \ ;$$
$$\frac{\partial u'}{\partial t'} + u'\frac{\partial u'}{\partial x'} + c'^2\frac{\partial(ln\ \rho')}{\partial x'} = 0 \qquad (5.1)$$

From Equations (D-7) and (D-18), $c'$ and $\rho'$ are related to $p'$ as

$$\rho' = p'^{1/\gamma}[1 + \Lambda_1(p')] \approx p'^{1/\gamma}exp[\Lambda_1(p')] \qquad (5.2)$$

$$\begin{aligned} c' =& ((\gamma-1)/2)^{1/2}\hat{Z}^{1/2}(p')^{(\gamma-1)/2\gamma}[1 + \Lambda_3(p')]^{1/2} \\ \approx& ((\gamma-1)/2)^{1/2}(p')^{(\gamma-1)/2\gamma}[1 + (1/2)(\hat{B} - \hat{A} + \Lambda_3(p'))] \\ \approx& ((\gamma-1)/2)^{1/2}(p')^{(\gamma-1)/2\gamma}exp[(1/2)(\hat{B} - \hat{A} + \Lambda_3(p'))] \qquad (5.3) \end{aligned}$$

Thus, dropping the approximation notation, it is found that

$$\begin{aligned} ln\ p' =& \gamma[ln\ \rho' - \Lambda_1(p')] \\ =& (2\gamma/(\gamma-1))[ln\ c' - (1/2)ln\ (\gamma-1)/2) \ - (1/2)(\hat{B} - \hat{A} + \Lambda_3(p'))] \qquad (5.4) \end{aligned}$$

Finally, $\rho'$ and $c'$ can be related; specifically,

$$ln\ \rho' = (2/(\gamma - 1))ln\ c' + \Lambda_1 - \Lambda_3/(\gamma - 1) \tag{5.5}$$

plus additive constants that are eliminated upon differentiation. The factor $\Lambda_1 - \Lambda_3/(\gamma-1)$ is the fractional variation of the isentropic real-gas relation between $c^2$ and $\rho$ from the isentropic ideal-gas relation between those variables.

## 5.1. *Riemann invariant construction*

Define $\epsilon \equiv \Lambda_1 - \Lambda_3/(\gamma - 1)$ and substitute into Equation (5.1) for $\rho'$.

$$\frac{2}{\gamma - 1}\left[\frac{\partial c'}{\partial t'} + u'\frac{\partial c'}{\partial x'}\right] + c'\frac{\partial \epsilon}{\partial t'} + u'c'\frac{\partial \epsilon}{\partial x'} + c'\frac{\partial u'}{\partial x'} = 0 \tag{5.6}$$

$$\frac{\partial u'}{\partial t'} + u'\frac{\partial u'}{\partial x'} + c'\frac{2}{\gamma - 1}\frac{\partial c'}{\partial x'} + c'^2\frac{\partial \epsilon}{\partial x'} = 0 \tag{5.7}$$

Using the lowest order relation $p' = (2c'^2/(\gamma - 1))^{\gamma/(\gamma-1)}$, $\epsilon, \Lambda_1$, and $\Lambda_3$ can be converted to functions of $c'$. Then, a function $\Psi(c') \equiv \int c' d\epsilon = \int c'(d\epsilon/dc')dc'$ is created. The function $\Psi$ provides for an isentropic path an integrated effect of the variation from the ideal-gas behavior. This allows re-organization of the continuity and momentum equations by addition and subtraction to yield

$$\left[\frac{\partial}{\partial t'} + (u' + c')\frac{\partial}{\partial x'}\right]\left(u' + \frac{2}{\gamma - 1}c' + \Psi(c')\right) = 0 \tag{5.8}$$

$$\left[\frac{\partial}{\partial t'} + (u' - c')\frac{\partial}{\partial x'}\right]\left(u' - \frac{2}{\gamma - 1}c' - \Psi(c')\right) = 0 \tag{5.9}$$

This yields the two modified Riemann Invariants

$$I_{\mathrm{R}} \equiv u' + \frac{2}{\gamma - 1}c' + \Psi(c') \ ;$$
$$I_{\mathrm{L}} \equiv u' - \frac{2}{\gamma - 1}c' - \Psi(c') \tag{5.10}$$

which propagate respectively along the characteristic paths

$$\frac{dx'}{dt'} = (u' + c') \ ;$$
$$\frac{dx'}{dt'} = (u' - c') \tag{5.11}$$

where the following relations are developed from the definitions given by Equations (D-8) and (D-16):

$$\begin{aligned}
\Lambda_1 \equiv\ & (\lambda_1-\lambda_{\rm b})\big[\big(\frac{p}{\hat{p}}\big)^{1/\gamma}-1\big]-\lambda_2\big[\big(\frac{p}{\hat{p}}\big)^{(2-\gamma)/\gamma}-1\big] \\
& +\lambda_3\big[\big(\frac{p}{\hat{p}}\big)^{(3-\gamma)/2\gamma}-1\big] \\
=\ & (\lambda_1-\lambda_{\rm b})\big[\big(\frac{2c'^2}{\gamma-1}\big)^{1/(\gamma-1)}-1\big]-\lambda_2\big[\big(\frac{2c'^2}{\gamma-1}\big)^{(2-\gamma)/(\gamma-1)}-1\big] \\
& +\lambda_3\big[\big(\frac{2c'^2}{\gamma-1}\big)^{(3-\gamma)/2(\gamma-1)}-1\big]
\end{aligned} \tag{5.12}$$

$$\begin{aligned}
\Lambda_3 \equiv\ & -\lambda_{\rm b}+\lambda_1-\lambda_2+\lambda_3+\frac{(\gamma+1)(\lambda_{\rm b}-\lambda_1)}{\gamma}\big(\frac{p}{\hat{p}}\big)^{1/\gamma} \\
& +\frac{2\lambda_2}{\gamma}\big(\frac{p}{\hat{p}}\big)^{(2-\gamma)/\gamma}-\frac{(3+\gamma)\lambda_3}{2\gamma}\big(\frac{p}{\hat{p}}\big)^{(3-\gamma)/2\gamma} \\
=\ & -\lambda_{\rm b}+\lambda_1-\lambda_2+\lambda_3+\frac{(\gamma+1)(\lambda b-\lambda_1)}{\gamma}\big(\frac{2c'^2}{\gamma-1}\big)^{1/(\gamma-1)} \\
& +\frac{2\lambda_2}{\gamma}\big(\frac{2c'^2}{\gamma-1}\big)^{(2-\gamma)/(\gamma-1)}-\frac{(3+\gamma)\lambda_3}{2\gamma}\big(\frac{2c'^2}{\gamma-1}\big)^{(3-\gamma)/2(\gamma-1)}
\end{aligned} \tag{5.13}$$

$$\begin{aligned}
\epsilon \equiv\ & \Lambda_1-\frac{\Lambda_3}{\gamma-1}=\frac{\gamma}{\gamma-1}(\lambda_{\rm b}-\lambda_1+\lambda_2-\lambda_3)+\frac{(\gamma^2+1)(\lambda_1-\lambda_{\rm b})}{\gamma(\gamma-1)}\big(\frac{2c'^2}{\gamma-1}\big)^{1/(\gamma-1)} \\
& -\frac{(\gamma^2-\gamma+2)\lambda_2}{\gamma(\gamma-1)}\big(\frac{2c'^2}{\gamma-1}\big)^{(2-\gamma)/(\gamma-1)} \\
& +\frac{(2\gamma^2-\gamma+3)\lambda_3}{2\gamma(\gamma-1)}\big(\frac{2c'^2}{\gamma-1}\big)^{(3-\gamma)/2(\gamma-1)}
\end{aligned} \tag{5.14}$$

Consequently,

$$\begin{aligned}
c'\frac{d\epsilon}{dc'} =\ & 2c'^2\frac{d\epsilon}{d(c'^2)} \\
=\ & 2\frac{(\gamma^2+1)(\lambda_1-\lambda_{\rm b})}{\gamma(\gamma-1)^2}\big(\frac{2c'^2}{\gamma-1}\big)^{1/(\gamma-1)}-2\frac{(\gamma^2-\gamma+2)(2-\gamma)\lambda_2}{\gamma(\gamma-1)^2}\big(\frac{2c'^2}{\gamma-1}\big)^{(2-\gamma)/(\gamma-1)} \\
& +\frac{(2\gamma^2-\gamma+3)(3-\gamma)\lambda_3}{2\gamma(\gamma-1)^2}\big(\frac{2c'^2}{\gamma-1}\big)^{(3-\gamma)/2(\gamma-1)}
\end{aligned} \tag{5.15}$$

$$\begin{aligned}
\Psi(c') =\ & \int c\frac{d\epsilon}{dc'}dc'=\frac{(\gamma^2+1)(\lambda_1-\lambda_{\rm b})}{\gamma(\gamma+1)}\big(\frac{2}{\gamma-1}\big)^{\gamma/(\gamma-1)}c'^{(\gamma+1)/(\gamma-1)} \\
& -\frac{(\gamma^2-\gamma+2)(2-\gamma)\lambda_2}{\gamma(3-\gamma)}\big(\frac{2}{\gamma-1}\big)^{1/(\gamma-1)}c'^{(3-\gamma)/(\gamma-1)} \\
& +\frac{(2\gamma^2-\gamma+3)(3-\gamma)\lambda_3}{8\gamma}\big(\frac{2}{\gamma-1}\big)^{(\gamma+1)/2(\gamma-1)}c'^{2/(\gamma-1)}
\end{aligned} \tag{5.16}$$

From Equations (5.10) and (5.11), it is concluded that $I_{\rm R}$ is a function of $\eta \equiv t'-\int(u'+c')^{-1}dx'$ alone and $I_{\rm L}$ is a function of $\xi \equiv t'-\int(u'-c')^{-1}dx'$ alone. Integrals are taken along the paths indicated by Equation (5.11). If waves travel in one direction only, $u'$ and $c'$ are constant along each characteristic path. For example, if waves travel only in the positive $x'$-direction, $\eta = t' - (x' - x'_{\rm P})/(u' + c')$ while, if waves travel only in the negative $x'$-direction, $\xi = t' - (x' - x'_{\rm P})/(u' - c')$. For the moment, consider $x'_{\rm P}$ as a reference position. Later in a specific example, that position is identified as the location of a moving boundary.

From Equation (5.10), it follows that

$$
\begin{aligned}
u' &= \frac{I_{\rm R}(\eta) + I_{\rm L}(\xi)}{2} \quad ; \\
c' &= \frac{\gamma - 1}{4}[I_{\rm R}(\eta) - I_{\rm L}(\xi) - 2\Psi(c')] \\
&\approx \frac{\gamma - 1}{4}\Big[I_{\rm R}(\eta) - I_{\rm L}(\xi) - 2\Psi(\frac{(\gamma - 1)(I_{\rm R}(\eta) - I_{\rm L}(\xi))}{4})\Big]
\end{aligned}
\tag{5.17}
$$

As an example of nonlinear but isentropic wave propagation, consider a sinusoidally oscillating piston at one end (i.e., left end near $x' = 0$) of a semi-infinite duct with constant cross-section. Before the rightward travelling wave arrives, the gas is quiescent at stagnation values with $u' = 0$ and, from Equation (D-18),

$$
c' = \sqrt{[(\gamma - 1)/2]\hat{Z}[1 + \Lambda_3(1)]}
\tag{5.18}
$$

At the moving piston face,

$$
\begin{aligned}
x'_{\rm P} &= -(U/(2\pi))\cos\omega t = -(U/(2\pi))\cos\omega' t' \quad ; \\
u'(t', x'_{\rm P}) &= U \sin\omega' t'
\end{aligned}
\tag{5.19}
$$

Here, $\omega' \equiv \omega L/\sqrt{2c_{\rm p}\hat{T}} = 2\pi\sqrt{\frac{\gamma-1}{2}}$, if the reference length $L$ equals the theoretical wavelength for propagation at frequency $\omega$ and ideal-gas acoustic wave speed for the stagnation conditions. This choice of reference value does not condition the actual wave speed or wavelength in this situation.

For this problem, $I_{\rm L} = -\sqrt{\frac{2}{\gamma-1}\hat{Z}[1 + \Lambda_3(1)]} - \Psi(\sqrt{\frac{\gamma-1}{2}})$ uniformly for all values of $\xi$. From the velocity boundary condition at the piston, it follows that, at $x = x_{\rm P}, I_{\rm R}(t') =$ $2U\sin\omega' t' + \sqrt{\frac{2}{\gamma-1}} + \Psi(\sqrt{\frac{\gamma-1}{2}})$. Thus, for all $x', t'$ values of interest

$$I_R(\eta) = 2U\sin\omega'\eta + \sqrt{\frac{2}{\gamma-1}\hat{Z}[1+\Lambda_3(1)]} + \Psi(\sqrt{\frac{\gamma-1}{2}}) \tag{5.20}$$

$$u'(x',t') = u'(\eta) = U\sin\omega'\eta \tag{5.21}$$

$$c'(x',t') = c'(\eta) = \frac{\gamma-1}{2}U\sin\omega'\eta + \sqrt{\frac{\gamma-1}{2}\hat{Z}[1+\Lambda_3(1)]} + \frac{\gamma-1}{2}\left[\Psi(\sqrt{\frac{\gamma-1}{2}}) - \Psi\left(\frac{\gamma-1}{2}U\sin\omega'\eta + \sqrt{\frac{\gamma-1}{2}}\right)\right] \tag{5.22}$$

$$u'+c' = \frac{\gamma+1}{2}U\sin\ \omega'\eta + \sqrt{\frac{\gamma-1}{2}\hat{Z}[1+\Lambda_3(1)]} + \frac{\gamma-1}{2}\left[\Psi(\sqrt{\frac{\gamma-1}{2}}) - \Psi\left(\frac{\gamma-1}{2}U\sin\ \omega'\eta + \sqrt{\frac{\gamma-1}{2}}\right)\right] \tag{5.23}$$

$$u'-c' = \frac{3-\gamma}{2}U\sin\ \omega'\eta - \sqrt{\frac{\gamma-1}{2}\hat{Z}[1+\Lambda_3(1)]} - \frac{\gamma-1}{2}\left[\Psi(\sqrt{\frac{\gamma-1}{2}}) - \Psi\left(\frac{\gamma-1}{2}U\sin\ \omega'\eta + \sqrt{\frac{\gamma-1}{2}}\right)\right] \tag{5.24}$$

In this problem of choice, the velocity $u'$ remains a simple sinusoidal function of the characteristic coordinate $\eta'$ while $c'$ is a more complicated function of that coordinate due to the real-gas correction. These functions are a consequence of the particular boundary condition and not a general rule. For example, if the wave described above reflected on the right at an open end of the duct or partially open end (e.g., orifice in a wall), the velocity in the reflecting wave would have a real-gas correction described through the $\Psi$ function. Another example would relate to disturbances of the type found in combustion instability problems where, at specific locations, the divergence of the velocity could be a function of the thermodynamic variables.

The difference in slopes of the characteristics between the positive and negative peaks of the wave gives a measure of compressive wave steepening and broadening of the expansion portion as the sinusoidal shape transforms towards an N-wave. Specifically,

$$\Delta(u'+c') = (\gamma+1)U + \frac{\gamma-1}{2}\left[\Psi\left(\sqrt{\frac{\gamma-1}{2}} - \frac{\gamma-1}{2}U\right) - \Psi\left(\sqrt{\frac{\gamma-1}{2}} + \frac{\gamma-1}{2}U\right)\right] \tag{5.25}$$

The real-gas properties modify the rate of steepening as indicated by the difference in the $\Psi$ function.

### 5.2. *Results for piston-driven wave*

Results are presented in figures 5, 6, and 7 for the same gases and stagnation conditions that were considered in the nozzle flow analysis. Non-dimensional velocity, sound speed and pressure versus non-dimensional spatial location is given for both the ideal-gas and real-gas calculations at a time $t' = 10$ which is roughly the time for a wave to travel four-to-six wavelengths. (The non-dimensional frequency increases with $\gamma$ here.) The non-dimensional velocity amplitude of the piston is given as $U = 0.02$ which is roughly fifteen-to-thirty times smaller than the speed of sound, depending on the temperature. With the normalization scheme, the results easily scale to any frequency of piston oscillation. Generally, the original sinusoidal waveform distorts in well-known fashion towards an $N$-shaped waveform; multi-valued solutions are allowed to develop in physical space to emphasis the wave distortion; of course, a shock discontinuity must form, leaving only single-valued solutions.

Velocity is shown in figure 5. The same amplitude is maintained and is determined by the piston-motion amplitude for all cases. The ideal-gas and real-gas solutions for $u(\eta)$ are identical sinusoidal functions, but solutions for $u(t', x')$ differ because the sound speeds differ. The real-gas wave generally moves slower than the ideal-gas wave. A modest exception occurs with nitrogen at the higher temperature. Ideal gases with higher $\gamma$ and lower molecular mass tend to move more wavelengths as expected. The real gas does not follow the same trend. The differences between real and ideal behavior is greater for carbon dioxide and argon than for nitrogen.

In figure 6, modest-to-profound differences in sound speed from the ideal-gas behavior are seen. At lower temperatures, especially for carbon dioxide, the average real-gas sound speed is lower than the average ideal-gas value. At higher temperatures, the nitrogen real-gas has a modestly higher sound speed. The real-gas sound speed amplitudes are generally higher than the ideal-gas amplitudes. The explanation follows from the facts that the $\Psi$ function is negative and it oscillates out-of-phase with the velocity and sound speed. Thus, $\Psi$ has a positive contribution in Equation (5.17), thereby increasing the value of $c'$ amplitude above the ideal-gas value.

Figure 7 shows that the real-gas pressure has the most profound differences from the ideal gas; pressure amplitude is substantially increased at all temperatures but especially for the lower-temperature carbon dioxide, where an approximate tripling and doubling of the ideal-gas pressure amplitude case occurs at certain temperatures. Through the Riemann invariant, the velocity $u$ and sound speed $c$ are related linearly for the ideal gas and with a weakly nonlinear relation for the real gas. Consequently, the magnitudes of their amplitudes are comparable. However, the relation between pressure and sound speed involves a power law with exponent $2\gamma/(\gamma-1)$ which affects the triatomic gas more than the diatomic gas (and in turn more than the monatomic gas). This power relation not only makes pressure amplitudes larger than other amplitudes for both the ideal and real gases. but also accentuates the difference between real and ideal gases. The higher pressure amplitude for the larger molecules is needed to achieve the sound-speed amplitude (or effectively the temperature and enthalpy amplitudes) demanded through the Riemann invariant by the piston-velocity magnitude; that is, the higher specific heat requires that the piston do more work raising the needed product of pressure and velocity.

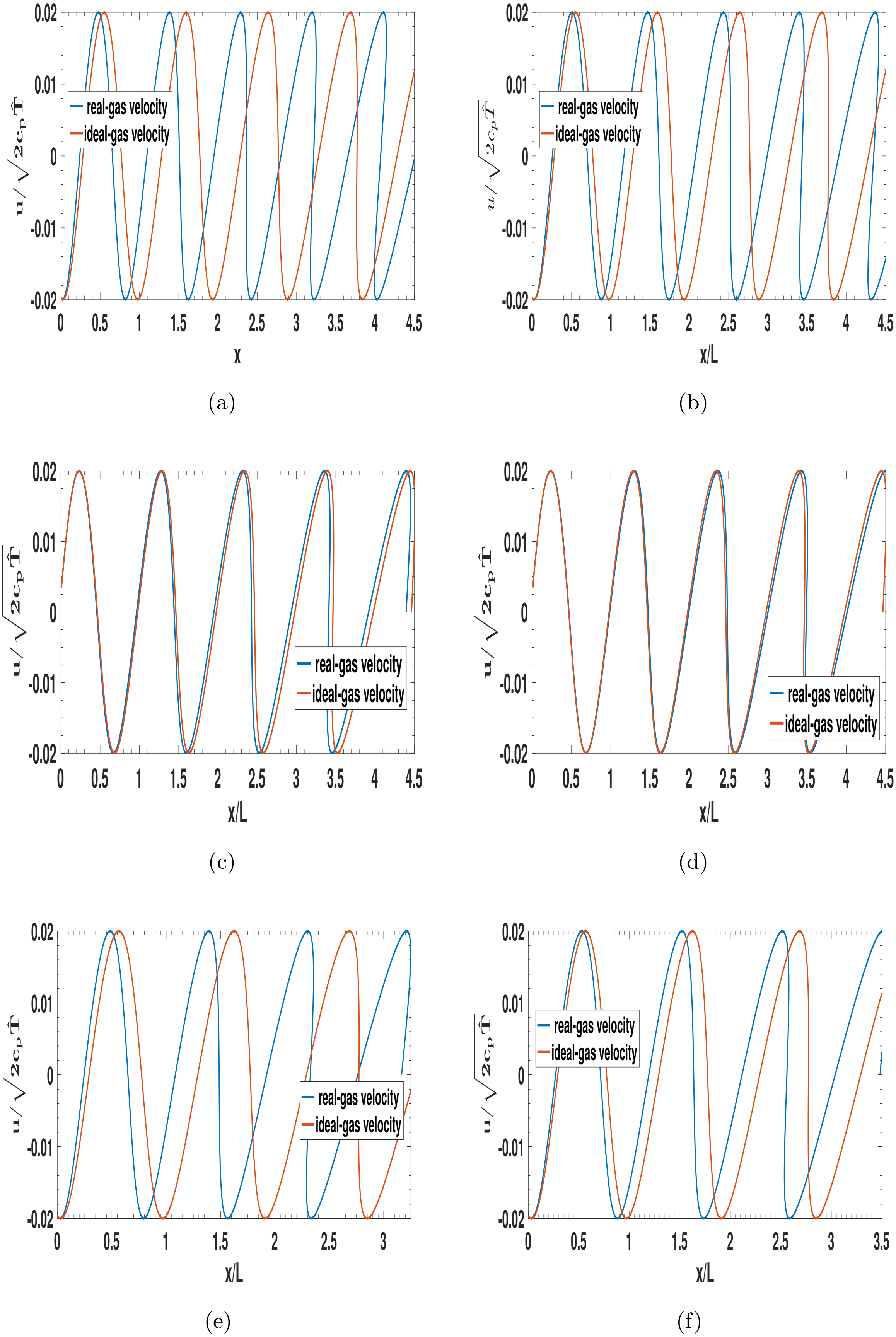

**Fig. 5 Comparison for piston-driven flow between real-gas flow and ideal-gas flow: non-dimensional velocity. (a) Argon, 300 K, 10 MPa; (b) Argon, 1000 K, 30 MPa; (c) Nitrogen, 400 K, 12 MPa; (d) Nitrogen, 1000 K, 30 MPa; (e) Carbon dioxide, 450 K, 10 MPa; (f) Carbon dioxide, 1000 K, 30 MPa.**

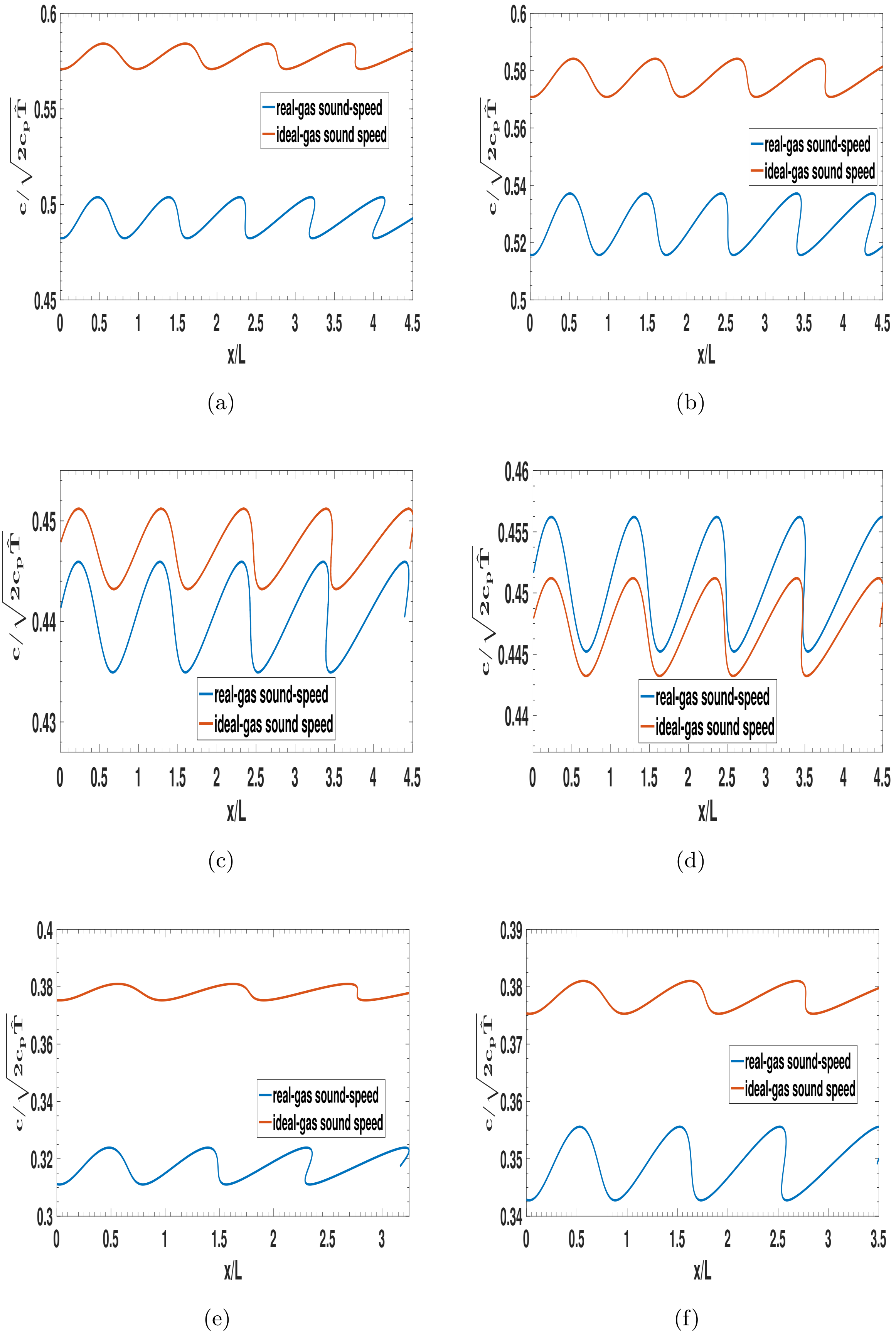

**Fig. 6 Comparison for piston-driven flow between real-gas flow and ideal-gas flow: non-dimensional sound speed. (a) Argon, 300 K, 10 MPa; (b) Argon, 1000 K, 30 MPa; (c) Nitrogen, 400 K, 12 MPa; (d) Nitrogen, 1000 K, 30 MPa; (e) Carbon dioxide, 450 K, 10 MPa; (f) Carbon dioxide, 1000 K, 30 MPa.**

The magnitudes of $Z-1$, which capture the difference between real and ideal gases, are significant. The compressibility factor is seen in Figure 7 to increase generally with temperature for these gases with values below one found for argon and carbon dioxide at lower temperatures. The amplitude of oscillation is more modest than found for other variables.

The author is unaware of previously published modifications to Riemann invariants owing to real-gas EoS effects.

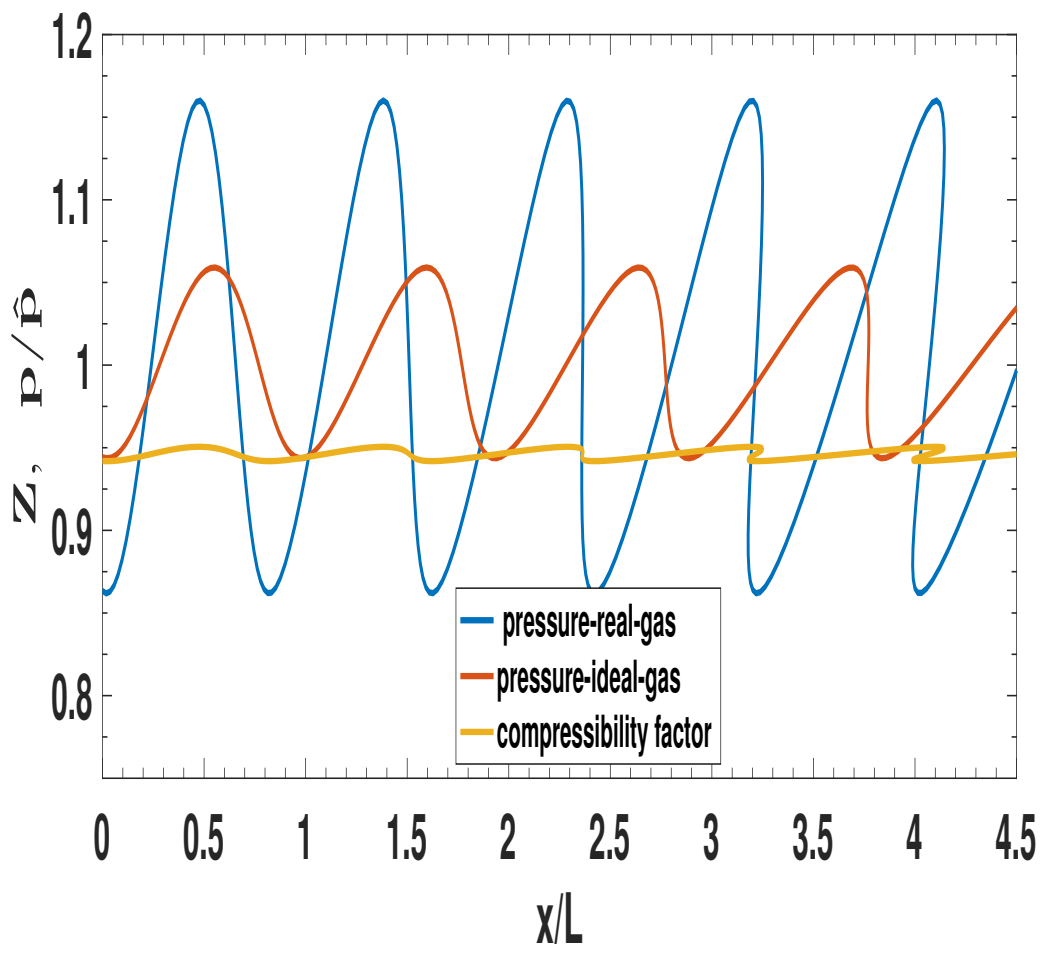

(a)

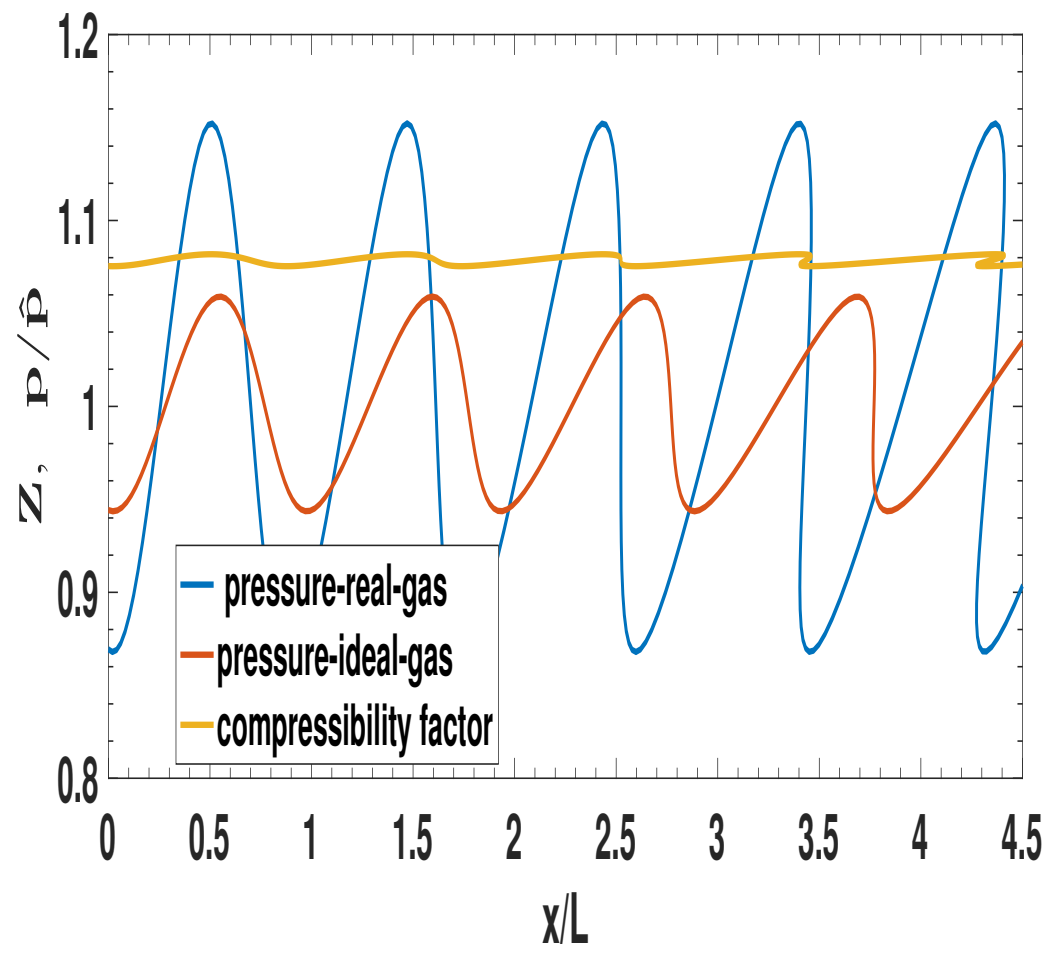

(b)

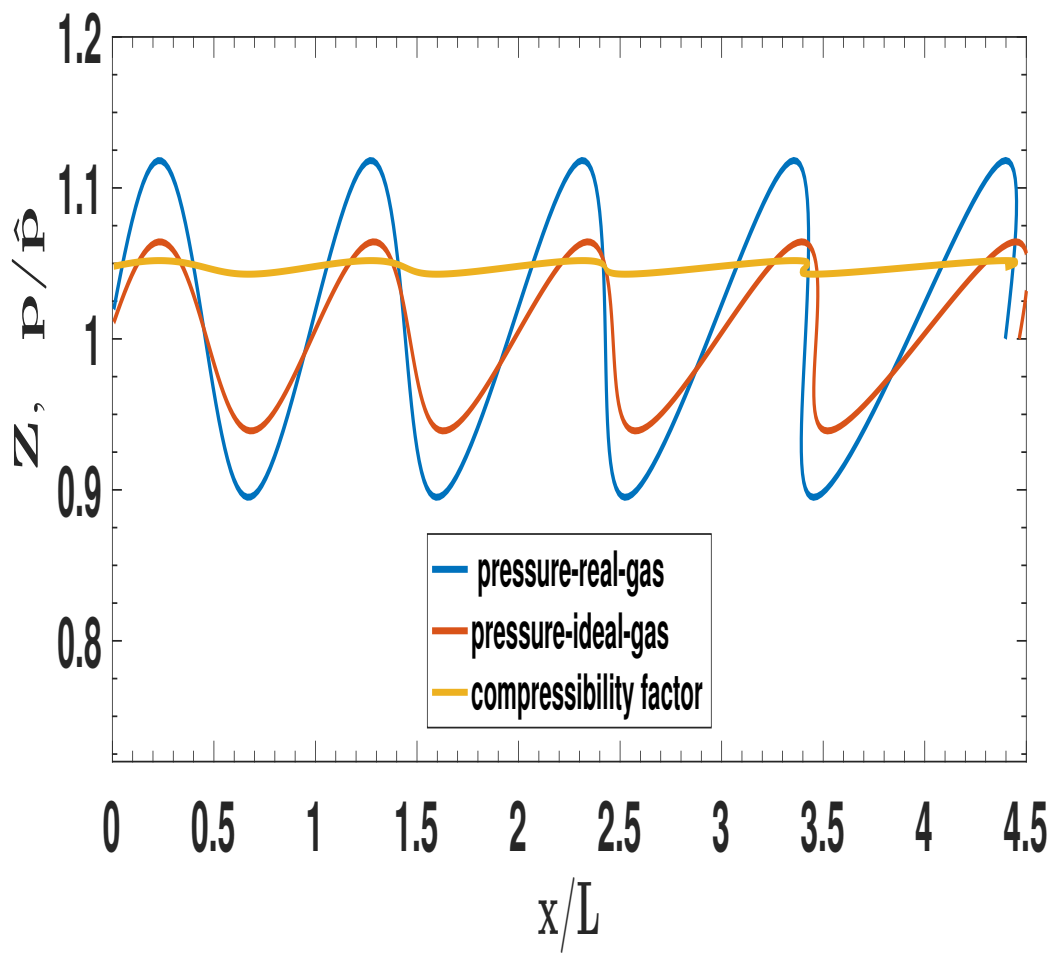

(c)

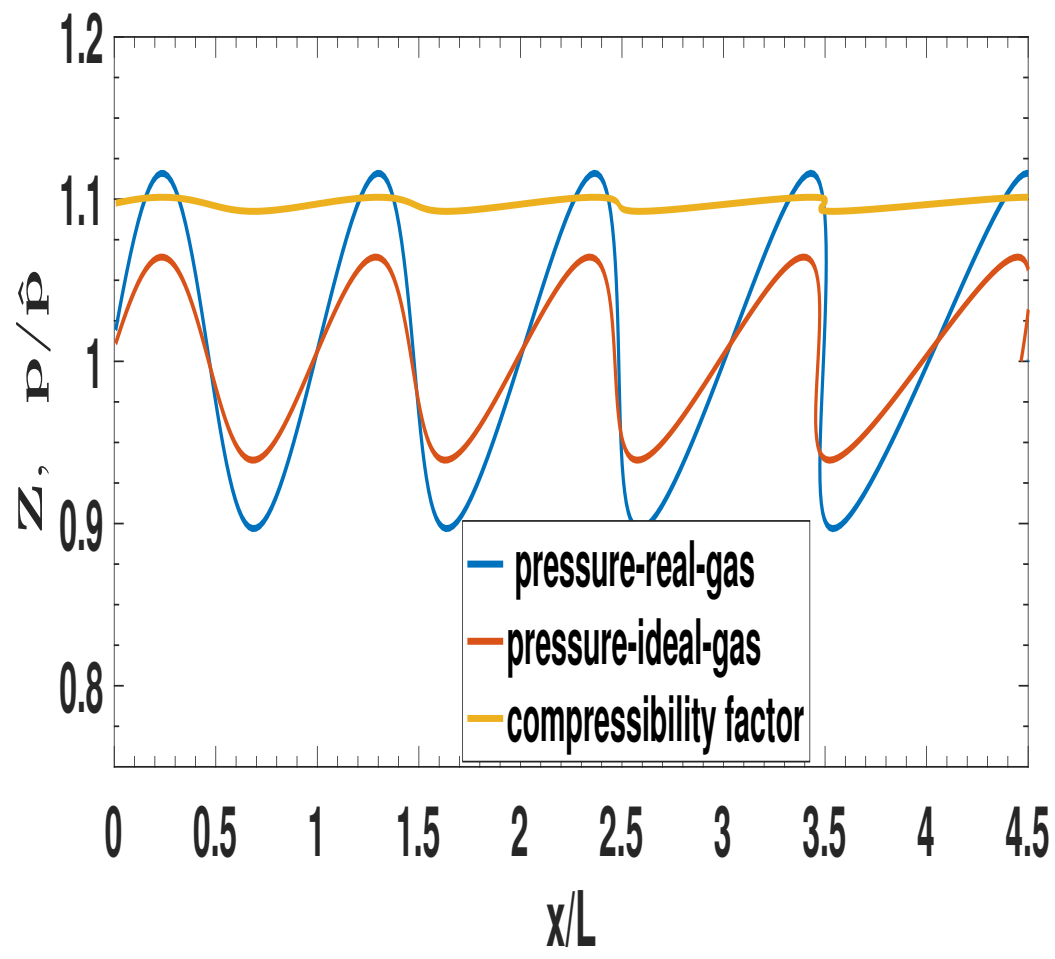

(d)

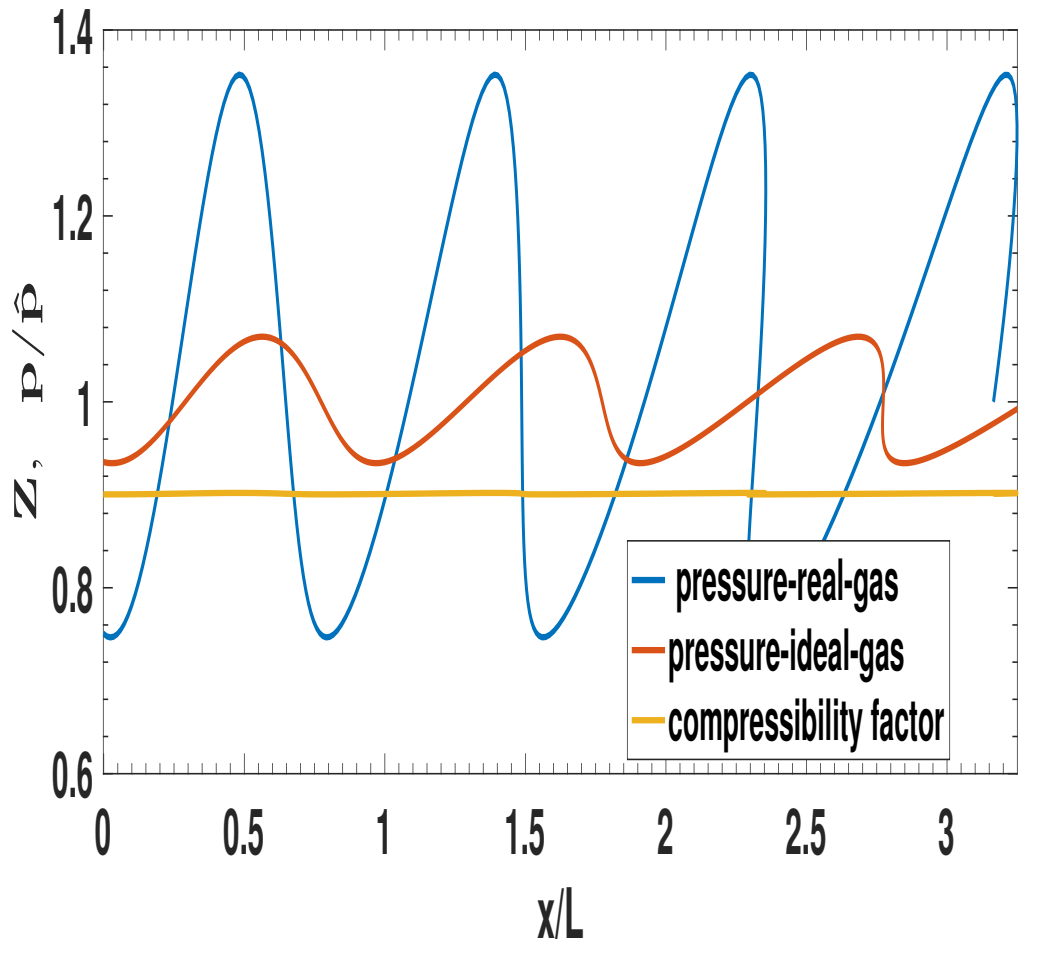

(e)

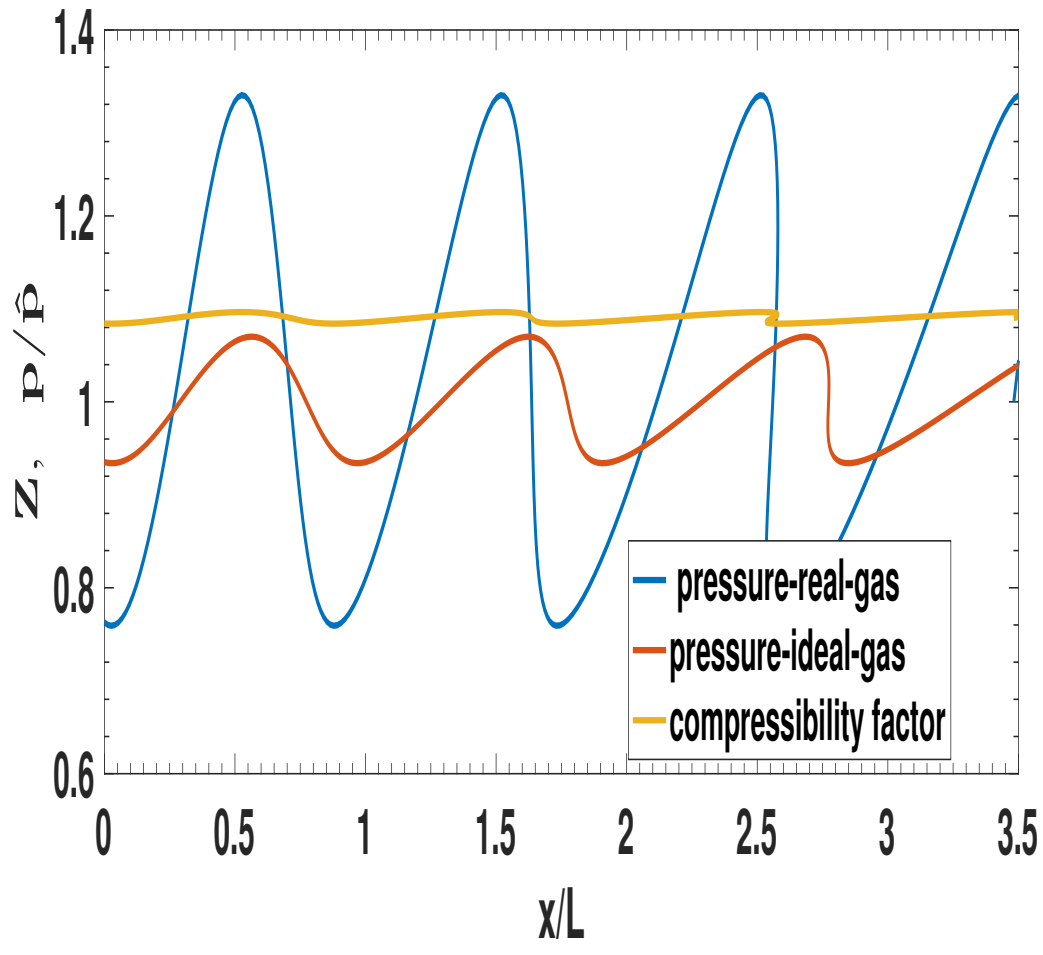

(f)

**Fig. 7 Comparison for piston-driven flow between real-gas flow and ideal-gas flow: non-dimensional pressure. (a) Argon, 300 K, 10 MPa; (b) Argon, 1000 K, 30 MPa; (c) Nitrogen, 400 K, 12 MPa; (d) Nitrogen, 1000 K, 30 MPa; (e) Carbon dioxide, 450 K, 10 MPa; (f) Carbon dioxide, 1000 K, 30 MPa.**

## 6. Shock Relations

The classical shock relations are built around the ideal-gas assumption, Specifically, the relation $h = \frac{\gamma}{\gamma-1}\frac{p}{\rho}$ is used. However, from Equation (B-5),

$$h = \frac{\gamma}{\gamma-1}\frac{p}{\rho Z} + RT\big[B - 2A + A'\big]$$
$$\approx \frac{\gamma}{\gamma-1}\frac{p}{\rho}\Big[1 + \frac{2-\gamma}{\gamma}A - \frac{1}{\gamma}B + \frac{\gamma-1}{\gamma}A'\Big] = \frac{\gamma}{\gamma-1}\frac{p}{\rho}\Big[1+\zeta\Big] \quad (6.1)$$

where $\zeta$ is defined by the last equation. The conservation equations for mass, normal momentum, transverse momentum, and energy across the shock wave can easily be manipulated into the following forms.

$$\frac{u_1}{u_2} = \frac{\rho_2}{\rho_1} \quad (6.2)$$
$$u_1 + \frac{p_1}{\rho_1 u_1} = u_2 + \frac{p_2}{\rho_2 u_2} \quad (6.3)$$
$$v_1 = v_2 \quad (6.4)$$
$$\frac{\gamma}{\gamma-1}\frac{p_1}{\rho_1}\Big[1+\zeta_1\Big] + \frac{u_1^2}{2} = \frac{\gamma}{\gamma-1}\frac{p_2}{\rho_2}\Big[1+\zeta_2\Big] + \frac{u_2^2}{2} = \frac{\gamma+1}{2(\gamma-1)}c^{*2} = \hat{h} - \frac{v_1^2}{2} \quad (6.5)$$

where $u$ and $v$ are the normal and transverse velocity components, respectively. Subscripts 1 and 2 pertain respectively to conditions on the upstream and downstream sides of the shock. Upstream conditions for $p_1, T_1, u_1$, and $v_1$ are regarded as given with these conservation laws determining the downstream values. The other interesting upstream variables can also be readily determined. Given $p_1$ and $T_1$, the value of $\rho_1$ is readily determined from $Z = 1 + B - A$. Equation (6.1) determines $\zeta_1$. The characteristic velocity $c^*$ is an abstract velocity with value between $u_1$ and $u_2$ that is determined by the stagnation enthalpy less the kinetic energy per unit mass associated with the transverse flow; so, with knowledge of $u_1$ and $v_1$, the values $\hat{h}_1 = \hat{h}_2$ and $c^* = \sqrt{[2(\gamma-1)/(\gamma+1)][\hat{h}_1 - v_1^2/2]}$ are determined. For the normal shock, $c^*$ is directly proportional to the square root of stagnation enthalpy.

### 6.1. *Modified Prandtl relation*

It is apparent that, for the ideal gas where $\zeta_1 = \zeta_2 = 0$, the enthalpy term in the energy conservation equation is easily related to the pressure term in the normal momentum conservation equation. However, a more complex relation exists for a real gas. Using the energy equation to substitute into the pressure term of the momentum equation, it

follows that

$$\left[\frac{\gamma+1}{\gamma-1}\frac{c^{*2}}{u_1}-u_1\right]\frac{1}{1+\zeta_1}+\frac{2\gamma}{\gamma-1}u_1=\left[\frac{\gamma+1}{\gamma-1}\frac{c^{*2}}{u_2}-u_2\right]\frac{1}{1+\zeta_2}+\frac{2\gamma}{\gamma-1}u_2 \tag{6.6}$$

After multiplying the left $c^{*2}$ by $u_2/u_2$ and the right $c^{*2}$ by $u_1/u_1$, expanding with the small values of $\zeta_1$ and $\zeta_2$, and factoring out $u_2 - u_1$, the relation becomes

$$1=\frac{c^{*2}}{u_1u_2}\left[1-\frac{\zeta_1u_2-\zeta_2u_1}{u_2-u_1}\right]-\frac{\gamma-1}{\gamma+1}\frac{\zeta_2u_2-\zeta_1u_1}{u_2-u_1} \tag{6.7}$$

To lowest order, the classical Prandtl relation $c^{*2} = u_1u_2$ is obtained and may be used to substitute into the higher order term yielding

$$\frac{c^{*2}}{u_1u_2}=1+\frac{\zeta_1u_2-\zeta_2u_1}{u_2-u_1}+\frac{\gamma-1}{\gamma+1}\frac{\zeta_2u_2-\zeta_1u_1}{u_2-u_1} \tag{6.8}$$

Consistently with the approximation here, $u_2 = c^{*2}/u_1$ may be substituted on the right side of this above equation to yield the modified Prandtl shock relation for high pressure environments.

$$\frac{c^{*2}}{u_1u_2}=1+\frac{\zeta_1c^{*2}-\zeta_2u_1^2}{c^{*2}-u_1^2}+\frac{\gamma-1}{\gamma+1}\frac{\zeta_2c^{*2}-\zeta_1u_1^2}{c^{*2}-u_1^2} \tag{6.9}$$

In some cases, numerical difficulties occur with the form of Equation (6.9) because some of the terms on the right side occasionally involve the ratio of two small numbers. It becomes more convenient at times to solve Equation (6.6) as a quadratic equation for $u_2$, using the negative-sign option in the classical formula and iterating to update volumes of $\zeta_2$.

For the limit of the ideal gas, the classical $c^{*2} = u_1u_2$ is recovered and, in the limit of shock strength going to zero for the ideal gas, $c^* = u_1 = u_2 = c$. This is not the general real-gas result; $c^{*2}$ remains proportional to the stagnation enthalpy but the relation with velocity is more complex.

The limiting characteristic velocity value can be determined as the strength of the shockwave goes to zero, i.e., $u_2 \to u_1$. Equations (6.1) and (6.9) are used with the definition of $c^*$ to obtain

$$c^{*2} = [1+\frac{2\gamma}{\gamma+1}\zeta_1]u_1^2 = 2\frac{\gamma-1}{\gamma+1}\hat{h}_1 = 2\frac{\gamma-1}{\gamma+1}[h_1+\frac{u_1^2}{2}] \tag{6.10}$$

$$\begin{aligned} u_1^2 &= \frac{(\gamma-1)h_1}{1+\gamma\zeta_1}\approx(\gamma-1)h_1(1-\gamma\zeta_1) \\ &= \frac{\gamma p_1}{\rho_1}(1+\zeta_1)(1-\gamma\zeta_1)\approx\frac{\gamma p_1}{\rho_1}[1-(\gamma-1)\zeta_1] \end{aligned} \tag{6.11}$$

This limiting velocity for the real gas is generally not the characteristic velocity. In particular, $c^*$ does not go to $c_1$ as shock strength goes to zero. $\zeta$ is the fractional departure of $(\gamma-1)h$ from $\gamma p/\rho$ and $-(\gamma-1)\zeta_1/2$ is the fractional departure of the limiting wave speed from $\sqrt{\gamma p_1/\rho_1}$. Thus, $[Z_1-1-(\gamma-1)\zeta_1]/2$ is the fractional departure of the limiting wave speed from $\sqrt{\gamma RT_1}$. When the attraction parameter $A_1$ becomes larger (smaller) than the repulsion parameter $B_1$, $Z_1-1$ becomes negative (positive) but $\zeta_1$ tends towards becoming positive (negative). The change in sign does not occur simultaneously for $\zeta$ and $Z-1$. Nevertheless, real-gas limiting wave speed tends to be larger (smaller) than the ideal-gas value when $B_1 > A_1$ ($B_1 < A_1$).

An interesting set of normal shock calculations for real-gas flow of air at upstream values $T_1 = 700$K; $p_1 = 1$, 4, and 50 MPa is given by Kouremonos & Antonopoulos (1989). They use the original Redlich-Kwong form of the EoS but qualitative differences are not expected with our SRK form. Results are reported for the range $1.2 \leqslant M_1 \leqslant 5.5$ . That paper makes no mention of a modified Prandtl relation or modified Rankine-Hugoniot relation.

### 6.2. *Modified Rankine-Hugoniot relation*

Next, the modifications of the Rankine-Hugoniot relation are examined by manipulation of the conservation laws of Equation (6.5). Combination of the normal-momentum and continuity relations yields

$$u_1^2 = \frac{p_2-p_1}{\rho_2-\rho_1}\frac{\rho_2}{\rho_1} \;\; ; \;\; u_2^2 = \frac{p_2-p_1}{\rho_2-\rho_1}\frac{\rho_1}{\rho_2} \tag{6.12}$$

Substitution for the velocity terms in the energy equation and multiplication by the factor $\rho_1/p_1$ gives a linear relation for the pressure ratio $p_2/p_1$ in terms of $\rho_2/\rho_1, \gamma, \zeta_1$ and $\zeta_2$. Solution of that linear relation yields the modified Rankine-Hugoniot relation.

$$\frac{p_2}{p_1} = \left[\frac{\rho_2}{\rho_1} - \frac{\gamma-1}{\gamma+1+2\gamma\zeta_1}\right]\left[1 - \frac{\gamma-1}{\gamma+1+2\gamma\zeta_2}\frac{\rho_2}{\rho_1}\right]^{-1} \tag{6.13}$$

Equations (6.9 ) and (6.13) give the classical Prandtl and Rankine-Hugoniot relations when $\zeta_1 = \zeta_2 = 0$. Solutions to the modified relations can readily be obtained in a two-step iteration. First, taking $\zeta_1 = \zeta_2 = 0$ and given upstream values for $u_1, v_1, p_1$, and $\rho_1$, Equations (6.9 ) and (6.13), together with transverse-momentum and continuity equations in Equation (6.5), yield the zeroeth-order approximations to the downstream values, i.e., $u_2^*, p_2^*$, and $\rho_2^*$ and the correct value for $v_2$. Next, using these values for $p_2^*$ and $\rho_2^*$ with the ideal-gas relation, the values of $T_2^*$ and $\zeta_2$ are determined with sufficient accuracy. Substitution of $\zeta_1$ and $\zeta_2$ into the Equations (6.9, 6.13) and the continuity equation gives $u_2, p_2$, and $\rho_2$ with the desired accuracy. Then, $T_2$ is determined from $Z = 1 + B - A$ and $h_2 = \hat{h} - (u_2^2 + v_2^2)/2$.

As the pressure ratio $p_2/p_1 \to \infty$ in Equation (6.13), it follows that

$$\frac{\rho_2}{\rho_1} \to \frac{\gamma + 1 + 2\gamma\zeta_2}{\gamma - 1}$$

$$\zeta_2 = \frac{2-\gamma}{\gamma}A - \frac{1}{\gamma}B + \frac{\gamma - 1}{\gamma}A' \to \frac{1}{\gamma}\left[\frac{\tilde{a}S^2}{R_u T_c} - b\right]\frac{p_2}{R_u T_2} = \frac{1}{\gamma}\left[\frac{\tilde{a}S^2}{R_u T_c} - b\right]\frac{\rho_2}{W} \quad (6.14)$$

In the limits for $A$ and $A'$, it has been considered that temperature ratio goes to infinity as pressure ratio goes to infinity. For the ideal gas, $\rho_2/\rho_1|_{ideal} \to (\gamma + 1)/(\gamma - 1)$, giving a finite limiting value for $\rho_2$. Thus, $\zeta_2$ has a finite limit, thereby yielding, for the real gas the finite limit

$$\left.\frac{\rho_2}{\rho_1}\right|_{real} \to \frac{\gamma + 1 + 2\left[\frac{\tilde{a}S^2}{R_u T_c} - b\right]\frac{\rho_2}{W}}{\gamma - 1} = \left(1 + 2\frac{\rho_1}{W(\gamma - 1)}\left[\frac{\tilde{a}S^2}{R_u T_c} - b\right]\right)\frac{\gamma + 1}{\gamma - 1} \quad (6.15)$$

Differentiation of $p_2$ given by Equation (6.13) with respect to $\rho_2$, holding upstream values constant, and taking the limit as $p_2 \to p_1$ yields the result

$$\frac{dp_2}{d\rho_2}\Big|_{p_2 \to p_1} = \frac{\gamma p_1}{\rho_1}\left[1 - (\gamma - 1)\zeta_1 - \frac{\gamma - 1}{\gamma + 1}\rho_1 \frac{d\zeta_2}{d\rho_2}\Big|_{p_2 \to p_1}\right] \quad (6.16)$$

The derivative $d\zeta_2/d\rho_2$ in Equation (6.16) is taken along the path $p_2(\rho_2)$ defined by Equation (6.13). Specifically, $\zeta_2$ taken from Equation (6.1) should be cast as $\zeta_2(p_2, \rho_2)$. The derivative of $\zeta_2$ involves both the explicit derivative and the implicit derivative through $p_2(\rho_2)$. For the latter derivative, the approximate form $dp_2/d\rho_2|_{p_2 \to p_1} = \gamma p_1/\rho_1$ suffices in this higher-order term. As known, for the ideal gas, in this limit of shock strength going to zero, the derivative along the Rankine-Hugoniot curve given by Equation (6.16) goes to the value for the derivative along the isentropic particle path in that limiting situation. Namely, $dp_2/d\rho_2|_{p_2 \to p_1} \to \gamma p_1/\rho_1 = c^2 = \partial p_2/\partial \rho_2|_s$. For the general case of the real gas, this tangency also occurs.

The entropy gain across the shock is of third order in non-dimensional pressure gain for an ideal gas. This means that, in the limiting behavior of a weak shock, $Tds = dh - (1/\rho)dp << dh \approx (1/\rho)dp$ along the direct integration path (monotonic variations) from upstream to downstream conditions. A second-order accurate measure of this condition can be created using the mean-value theorem. Namely, the magnitude of $\mathit{\Delta} \equiv (\rho_1/p_1)[h_2-$ $h_1] - 2[p_2/p_1 - 1]/[\rho_2/\rho_1 + 1]$ can be compared to the magnitude of $(\rho_1/p_1)[h_2 - h_1]$ (or $2[p_2/p_1 - 1]/[\rho_2/\rho_1 + 1]$).

### 6.3. *Shock results*

Results for two cases with nitrogen gas are examined in figures 8 and 9, with upstream flow values for temperature and pressure given by 400 K, 10 MPa and 300 K, 3 MPa, respectively. These examples involve upstream conditions at supercritical pressure and supercritical temperature and at subcritical pressure and supercritical temperature, respectively. An attempt is made to choose upstream values that keep errors due to linearization small in the downstream flow. Among other things, this disallows treatment of compressible liquids at supercritical pressures and subcritical temperatures. Nitrogen is favored because it has the lowest critical values of the gases selected here for computations in other sections; thereby the upstream pressure and temperature are taken at sufficiently low values to keep the $A$ and $B$ parameters behind the shock low enough to validate the linearization.

Calculations are made over a range of $u_1$ values and displayed in the figures. Some portions of the range are not physically reasonable since the Second Law is not reflected in the algorithms. For example, portions of the curves where ratios of pressure, density, temperature, and enthalpy drop below values of one have been disregarded and are not shown in the figures.

Significant differences in the Rankine-Hugoniot (R-H) plots are generally seen in sub-figures 8a and 9a. The largest differences in pressure ratio between the real gas and the ideal gas occur near the limiting density ratios which themselves differ substantially. In these cases, the real gas has a smaller value for the upper limit on density ratio. Both the real and ideal cases are each calculated two ways as an error estimate on the linearization: (i) downstream pressure and density are calculated and then the ratios are "directly" formed; and (ii) the R-H formula is calculated. The error is small enough to make useful conclusions. These sub-figures and other results not shown here indicate that the real-gas pressure ratio generally appears larger (smaller) than the ideal-gas ratio when $B > A (A > B)$. (This should not be taken as a strict rule since quantities such as $A'$ and $A''$ can have influence.) At some values of density ratio, the R-H results from sub-figures 8a and 9a show very large differences in pressure ratio between the ideal gas and the real gas; in particular, the real-gas shock is much stronger there. This behavior is consistent with the results for the continuous wave given in Figure 7 where real gases had larger pressure amplitudes for the same forcing mechanism; the continuous waves there are expected to deform to N-shaped waveforms with shock formation. Note however that, at the same shock velocity $u_1$, the ideal gas can yield the greater pressure ratio as shown in sub-figures 8b and 9b. The velocity ratio however will be given as the reciprocal of the density ratio; thus the fractional change in velocity is smaller for the real gas in these cases.

Enthalpy and temperature show differences for the real gas in sub-figures 8c and 9c; downstream ideal-gas temperature exceeds real-gas temperature for the same shock velocity. Generally, non-dimensional enthalpy exceeds non-dimensional temperature for the real gas. Sub-figures 8d and 9d show that, for the given upstream conditions, the shock Mach number is smaller for the real gas. The sub-figures 8e and 9e show that a portion of domain has the values of $A$ and $B$ within desirable constraints for accuracy; however, for other portions, they achieve magnitudes near 0.4 which raises our error estimates to above 10%.

Figure 8f shows the results for nitrogen at upstream values of $400K, 10MPa$. In similar fashion to the previous example in the figure, an inflow velocity of about 450 m/s, $p_2 \to p_1, \rho_2 \to \rho_1, h_2 \to h_1, T_2 \to T_1$, and $Z_2 \to Z_1$ are found. The figure also shows that the approximate measure $\Delta$ related to entropy change is going to the zero limit and is higher order in magnitude.

The low-temperature nitrogen case is examined through figure 9. Here, the limiting behavior presents no surprises. At an inflow velocity of 357 m/s, $p_2 \to p_1, \rho_2 \to \rho_1, h_2 \to h_1, T_2 \to T_1$, and $Z_2 \to Z_1, M_2 \to M_1 \to 1$. Within our error here, $Z_1 = 1$ ; the limit should show ideal-gas behavior. In the calculations here (including unpublished cases), no cases with $M_1 < 1$ and entropy gain were found; they should be physically unstable if they exist as mathematical solutions.

Some analytical support can be given for the finding of limiting velocities not at the sonic speed. The relation between enthalpy and sound from Equations (C-16) and (6.1)

yields

$$\gamma \frac{p}{\rho} = \frac{c^2}{1+\sigma} = \frac{(\gamma-1)h}{1+\zeta} \tag{6.17}$$

Thus,

$$h \approx \frac{c^2}{\gamma-1}[1+\zeta-\sigma] \tag{6.18}$$

The energy conservation across the shock may be developed as follows and combined with the momentum relation.

$$\begin{aligned} h_2 - h_1 &= \frac{u_1^2 - u_2^2}{2} = \frac{u_1+u_2}{2}(u_1-u_2) \ ; \\ \delta h &= \left.\frac{u_1+u_2}{2\rho_1 u_1}\right|_{u_2 \to u_1} \delta p \to \frac{1}{\rho_1}\delta p \end{aligned} \tag{6.19}$$

The definitions $\delta h = h_2 - h_1, \delta u = u_2 - u_1$, etc. are applied in the limit as the jump across the shock is disappearing. As the jump across the shock becomes small, the asymptote is giving an isentropic result as shown by comparison with the differential relation that describes the combined First and Second Law, i.e., $Tds = dh - (1/\rho)dp$.

Table 4 compares present approximate calculations with cubic-equation computations of Kouremonos & Antonopoulos (1989), now designated as KA. Ratios of pressure and temperature plus downstream Mach number are compared for certain upstream Mach numbers. Subscripts $KA$ and $S$ are used in the table for the results of Kouremonos & Antonopoulos (1989) and the current results, respectively. The KA computations were done for a normal shock in air using the Redlich-Kwong EoS while the S results treat a normal shock in nitrogen and use the linearized SRK EoS. The quantitative KA results were interpreted from the graph in Figure 2 of their paper; so, the number of trusted significant digits was limited. They made no comparison with ideal-gas results. Table 4 shows that KA and S results compare favorably. For the chosen range of $M_1$, $Z_2$ varied from 1.02 to 1.06 in the S results, increasing with $M_1$; and the downstream pressure, temperature, and density were each lower than the value yielded for the ideal gas, with the difference increasing with $M_1$. The KA article also had results for 50 MPA which yields too high a value of $Z_2 - 1$ to apply our linearization and make a useful comparison; $Z_2 = 1.2$ and higher downstream.

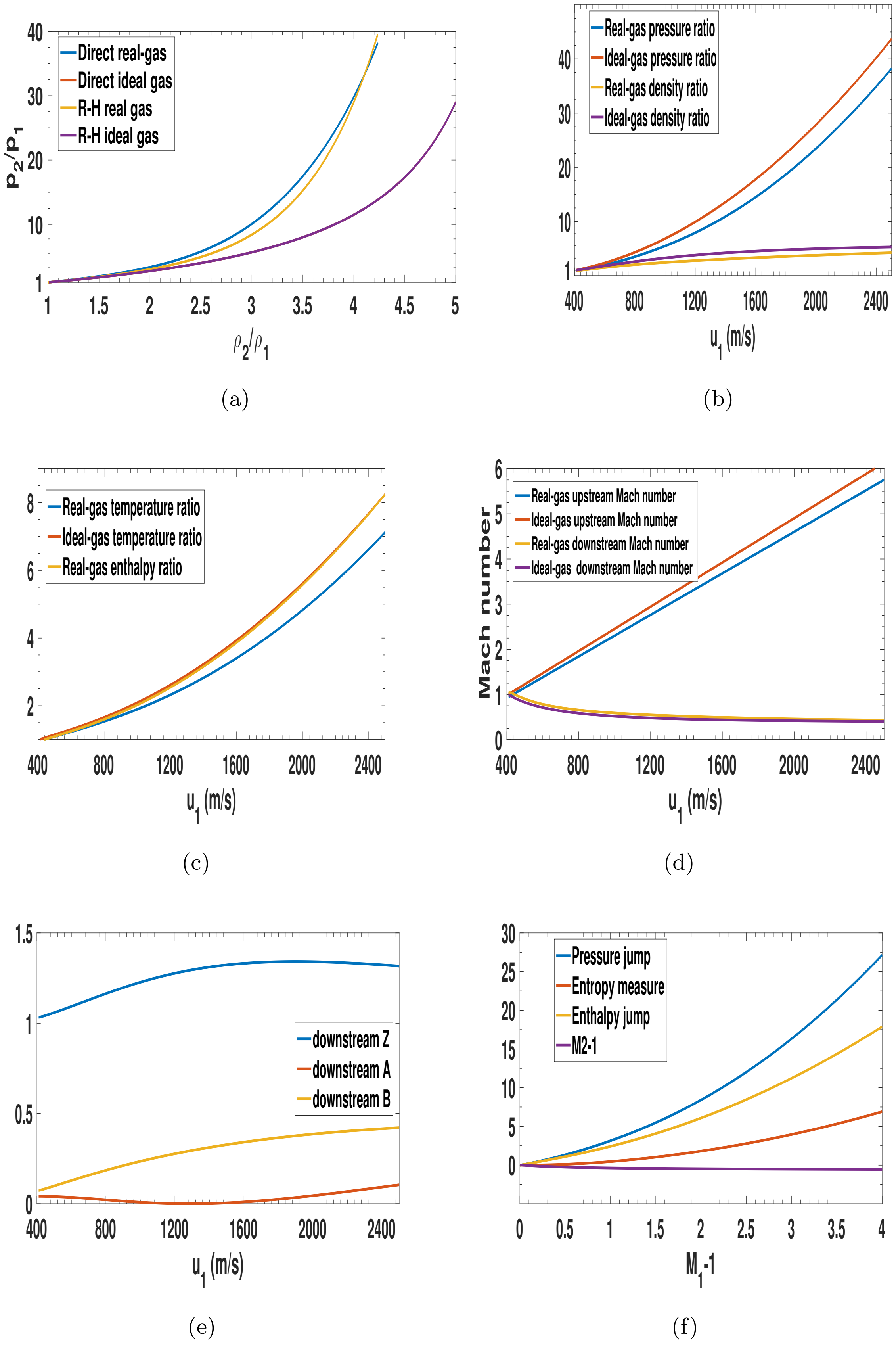

**Fig. 8 Shockwave: comparison of non-dimensional solutions between real gas and ideal gas for nitrogen; $T_1$ = 400 K, $p_1$ = 10 MPa, $u_1$ = 400-2500 m/s. (a) Rankine-Hugoniot relation; (b) Pressure ratio vs. shock velocity; (c) Enthalpy and temperature ratios; (d) Upstream and downstream Mach numbers; (e) Compressibility factor $Z$, attraction parameter $A$, repulsion parameter $B$; (f) Non-dimensional shock jumps in pressure, entropy, and enthalpy and $M_2 - 1$.**

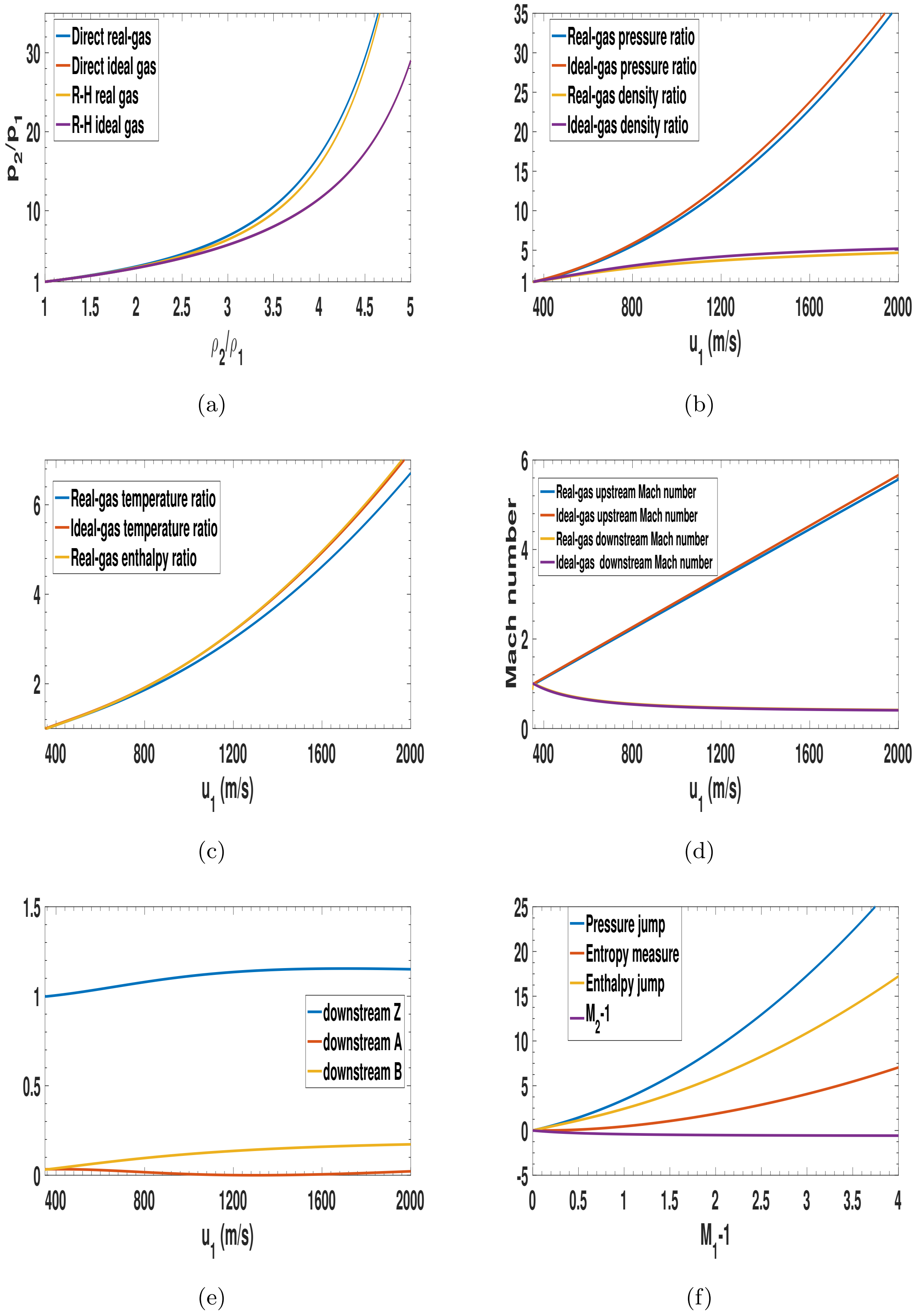

**Fig. 9 Shockwave: comparison of non-dimensional solutions between real gas and ideal gas for nitrogen; $T_1$ = 300 K, $p_1$ = 3 MPa, $u_1$ = 350-2000 m/s. (a) Rankine-Hugoniot relation; (b) Pressure ratio vs. shock velocity; (c) Enthalpy and temperature ratios; (d) Upstream and downstream Mach numbers; (e) Compressibility factor $Z$, attraction parameter $A$, repulsion parameter $B$; (f) Non-dimensional shock jumps in pressure, entropy, and enthalpy and $M_2 - 1$.**

Table 4: **Comparison with Kouremonos & Antonopoulos (1989) normal shock calculations. Values for upstream Mach number, downstream Mach number, temperature ratio, and pressure ratio. Upstream values were 700 K and 4 MPa.**

| $M_1$ | $M_{2KA}$ | $M_{2S}$ | $(T_1/T_2)_{KA}$ | $(T_1/T_2)_S$ | $(p_1/p_2)_{KA}$ | $(p_1/p_2)_S$ |
|---|---|---|---|---|---|---|
| 1.5 | 0.70 | 0.702 | 0.77 | 0.758 | 0.40 | 0.406 |
| 2.0 | 0.58 | 0.579 | 0.60 | 0.593 | 0.23 | 0.222 |
| 2.5 | 0.52 | 0.515 | 0.49 | 0.469 | 0.15 | 0.140 |
| 3.0 | 0.47 | 0.477 | 0.39 | 0.374 | 0.10 | 0.0963 |
| 3.5 | 0.45 | 0.453 | 0.31 | 0.302 | 0.07 | 0.0703 |
| 4.0 | 0.44 | 0.437 | 0.26 | 0.248 | 0.06 | 0.0538 |

## 7. Concluding Remarks

A method of linearization in parameter space has been shown to be useful in describing and explaining nonlinear real-gas behavior. The countering effects of intermolecular repulsion and attraction become more clearly visible. Monatomic, diatomic, and triatomic gases were studied at high and low temperatures. Generally, repulsion becomes more dominant at higher temperatures while attraction tends to prevail at lower temperatures. The method provides an accurate numerical description over a wide operating range for interesting compressible flows at elevated pressures. It is important to linearize the equation of state for enthalpy as well as the cubic equation of state for density; also, the speed-of-sound function must be properly expanded. The treatment identifies the substantial simplification of the ideal gas where enthalpy, sound-speed squared, temperature, and pressure-density ratio are all directly proportional to each other.

While the Soave-Redlich-Kwong cubic EoS has been chosen and single-component gases have been examined, the method for extension has been identified. Other well known cubic equations provide the same linear form with modest changes in parameter dependence on temperature. The rules for treating mixtures are identified in the literature and have been summarized here.

Three types of simple compressible flows have been treated: choked nozzle flow with expansion to supersonic flow, a nonlinear acoustical wave driven by an oscillating piston, and a normal shock wave. The differences amongst monatomic species, diatomic species, and triatomic species are often consequential. Interesting corrections to ideal-gas behavior are identified. Often, the corrections have different signs at high and low temperatures because of differences of relative strengths of the repulsion and attraction parameters (i.e., increases or decreases from the ideal-gas values). Corrections are found in the choked-nozzle discharge, optimal thrust, Riemann invariants, Prandtl shock relation, and Rankine-Hugoniot relation. Specifically, a study is made of the effects of variations from the three independent constants formed in ideal-gas treatment by the powerful relations $c^2 = (\gamma - 1)h = \gamma p/\rho = \gamma RT$. None of these equalities hold for the real gas.

Nozzle discharge coefficients could be greater or less than the ideal-gas value, depending on stagnation conditions and the particular gas. The different behaviors are related to the relative strengths of the attraction parameter $A$ and the repulsion parameter $B$ in the equation of state. No clear trends were seen for optimal thrust values.

A modified Rankine-Hugoniot relation and a modified Prandtl relation are developed for the real gas. Large differences in pressure ratio for the real and ideal gases are found near the limiting density ratio. As shock strength goes to zero for the real gas, the limiting speed is the sonic speed limit found also for the ideal gas.

The pressure amplitude in a piston-driven oscillation could be very large for the real gas, especially for a triatomic species. This behavior is consistent with results from the modified Rankine-Hugoniot results whereby pressure jumps for real-gas shocks can be substantially larger than jumps for the ideal-gas shocks. At lower temperatures, the real gas has a significantly lower sound speed than the ideal gas.

This research was supported by the National Science Foundation under Grant CBET-1333605 and by the Air Force Office of Scientific Research under Grant FA9550-15-1-0033.

Discussions with Professor Feng Liu about the steady, one-dimensional nozzle flow have been helpful. Editorial advice from Professor Said Elghobashi is valued.

## Appendix A: Comparison of Linear Results

Figure 10 compares for argon, nitrogen, and carbon dioxide the exact cubic solutions for $Z$ to the linear solutions for $Z$ for a few selected cases for temperature and pressure. A low-temperature case and a high-temperature case are taken for each gas, since it affects the magnitude of $Z-1$, sometimes even producing a change in sign. The linear solution is built around the smallness of $A$ and $B$, each of which increases with increasing pressure and decreasing temperature. Let us arbitrarily only accept an error in $Z$, if it is less than one per cent. The figure plots both the cubic relation $G(Z) = Z^3 - Z^2 + (A - B - B^2)Z - AB$ and the linear relation $H(Z) = Z - 1 - B + A$. $I(Z)$, the curve for second-order theory, is also plotted in figure 10 and are discussed below. The horizontal line gives the zero value so that the intersections with that line give $G(Z) = 0$ and $H(Z) = 0$. These intersections identify the solutions for $Z$. We see in sub-figures 10a,b that acceptable linear approximations for argon are found at $T$=300 K, $p$ = 10 MPa and $T$ =1000 K, $p$ = 30 MPa. Sub-figures 10 c,d, e,f show similar results for nitrogen are found at $T$=400 K, $p$ = 12 MPa and $T$ =1000 K, $p$ = 30 MPa and for carbon dioxide at $T$=450 K, $p$ = 10 MPa =1000 K, $p$ = 30 MPa. $A$ and $B$ each increase with increasing pressure and decrease with increasing temperature. Thus, these parameters can remain sufficiently bounded for our purpose here if temperature increases as pressure increases in a certain way.

The approximation concept can be extended to a polynomial solution with powers of $A$ and $B$ to make the approximation error as small as desired. For example, $Z = 1 + B - A - A^2 + 3AB$ with error of $O(A^3, A^2B, AB^2, B^3)$ can be used to approximate the solution to Equation (2.1). In figure 10, the $H(Z)$ and $I(Z)$ essentially give identical results. Figure 11 plots the function $I(Z) = Z - 1 + A - B + A^2 - 3AB$ along with functions $G(Z)$ and $H(Z)$ for the case where $T$ = 300 K and $p$ =20 MPa. The error for the linear approximation becomes unacceptable at this combination of a very high pressure and low temperature. An acceptable result emerges, however, for the second-order solution. The simplicity of the linear relation with error of $O(A^2, AB, B^2)$ is preferred in developing the flow solutions and further analysis is confined to domains where that error is very small. The second-order result here nevertheless demonstrates that there exists (i) a rational approximation method and (ii) a path to improvement for the temperature-pressure domains where the linear approximation is weak.

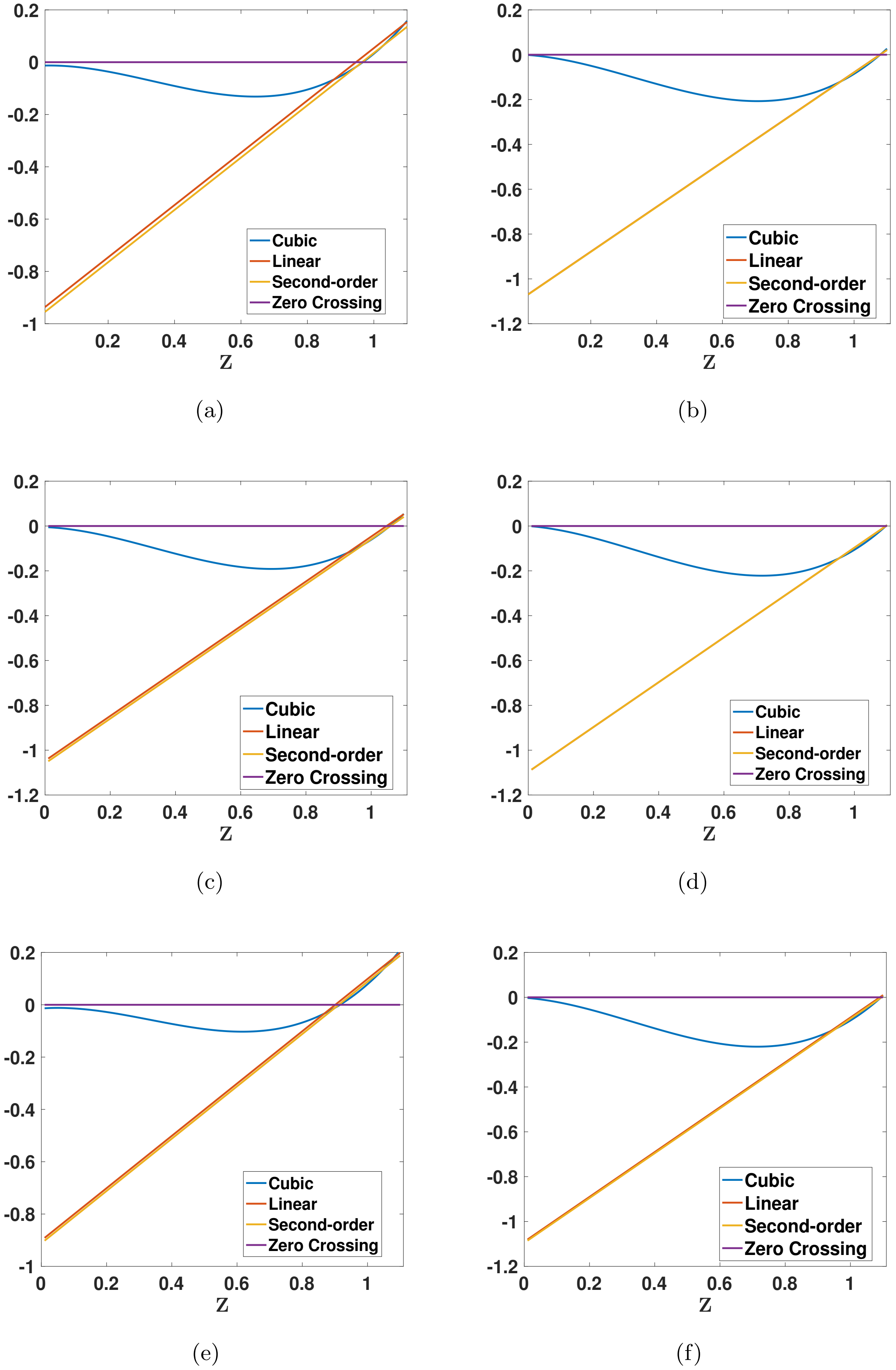

**Fig. 10 Sample comparisons of exact solution to cubic equation of state for argon, nitrogen, and carbon dioxide with local linear approximation. (a) Argon, 300 K, 10 MPa; (b) Argon, 1000 K, 30 MPa; (c) Nitrogen, 400 K, 12 MPa; (d) Nitrogen, 1000 K, 30 MPa; (e) Carbon dioxide, 450 K, 10 MPa; (f) Carbon dioxide, 1000 K, 30 MPa.**

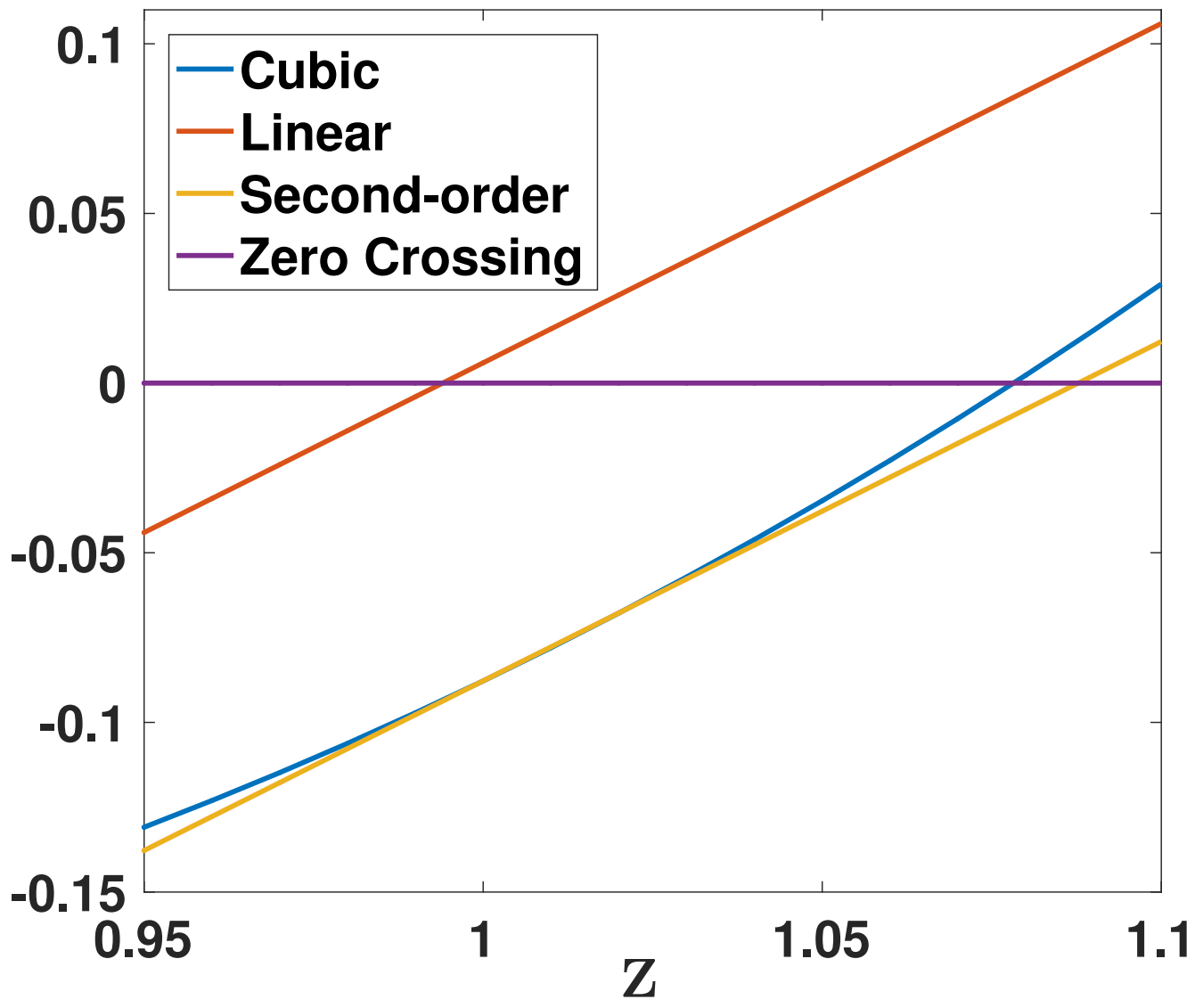

**Fig. 11 Comparison of exact solution, linear approximation, and second-order approximation for cubic equation for nitrogen at 300 K , 20 MPa.**

In order to ensure that second-order terms do not become too large, the linearization should be used where both $A$ and $B$ are O$(10^{-1})$. It is possible that $B - A$ is small in magnitude but $A$ and $B$ are individually too large for accurate use of the linear method. Figure 12 shows the range for certain magnitudes of those parameters for $CO_2$. Between the two lines in either sub-figure $0.01 < A < 0.10$ or $0.01 < B < 0.10$. There is a reasonably large range that covers interesting situations. To the left of both lines, the parameter is smaller than 0.01 while to the right of both lines, it is greater than 0.10. Qualitatively similar results are found for other gases.

The behavior of compressible flows at elevated pressure has qualitative differences that depend on whether the compressibility factor $Z$ is greater than or less than unity, or almost equivalently whether for a given pressure and temperature the density is less than or greater than the ideal-gas value. Figure 13 shows examples where Z is considered for isentropic expansion from given stagnation conditions over a range of pressure that varies by two orders of magnitude. Here, the temperature value is related to pressure through an isentropic relation. One can consider the values of $A, B, Z$ here to represent values found during an isentropic expansion (right-to-left in figure) or compression (left-to-right in the figure). At higher stagnation temperatures, $Z > 1$ is typical; repulsion (through parameter $B$) tends to be stronger than attraction (through parameter $A$) in determining the variation from an ideal-gas behavior. For stagnation temperature closer to (but still above) the critical temperature, $Z < 1$ often occurs; attraction becomes stronger than repulsion. So, in sub-figures 13b, d, f, $B > A, Z > 1$, and density are less than the ideal-gas value (except at the low-pressure end of the expansion (compression). On the contrary for sub-figures 13a,e, $A > B, Z < 1$, and density exceeds the ideal-gas value. Sub-figure 13c for lower temperature with nitrogen shows a transition between the fore-mentioned two regimes; $Z - 1$ changes value during the expansion (compression). These density values have consequence for mass flux and momentum flux in choked flows.

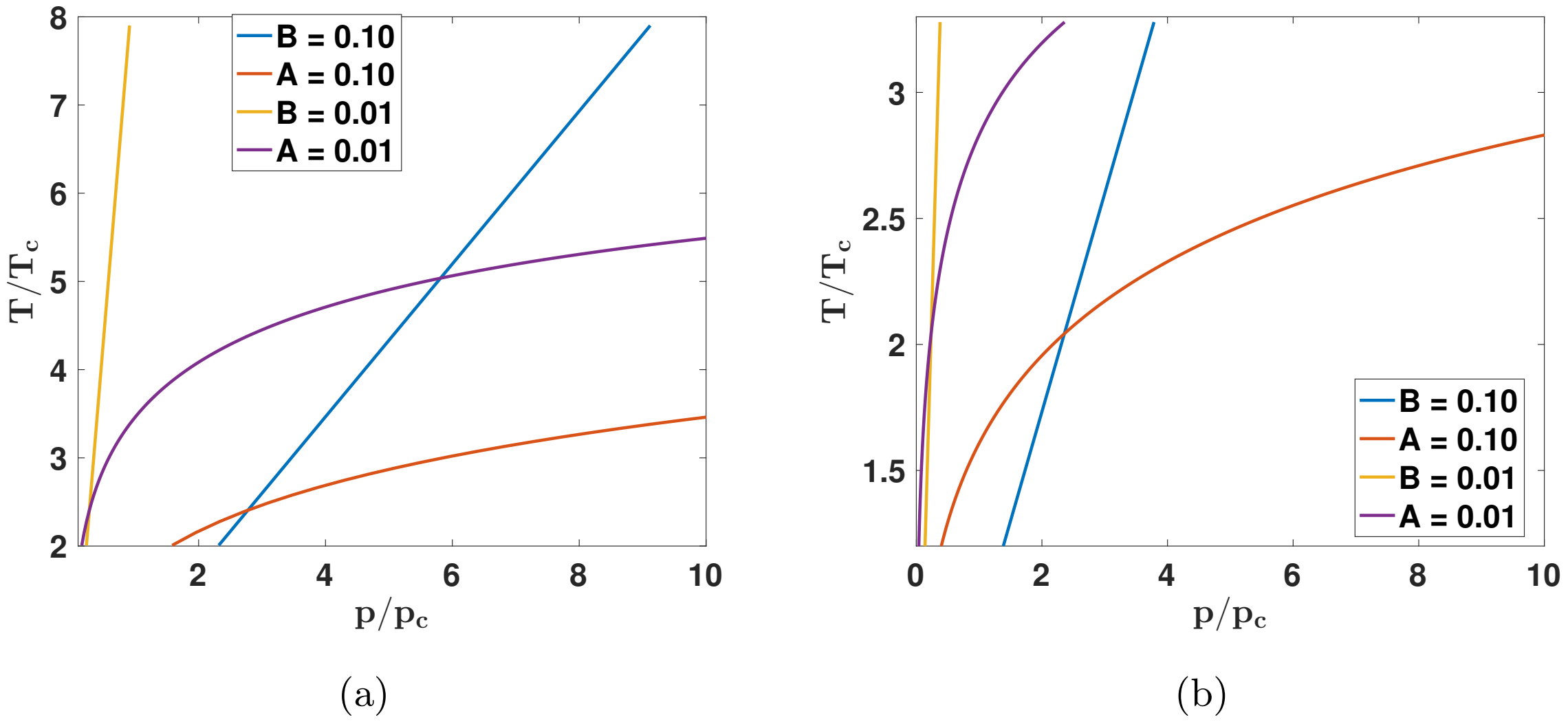

**Fig. 12 Range of pressure and temperature (normalized by critical values) where linear approximation can be useful for nitrogen and carbon dioxide. Parameter bounds: (a) Nitrogen, (b) Carbon dioxide.**

In Appendices B, C, and D, it is shown that other thermodynamic variables are analytic functions of the parameters and can be expanded in powers of $A$ and $B$; therefore, it can be expected that the linear approximations for those variables have the same error bounds as $Z - 1$.

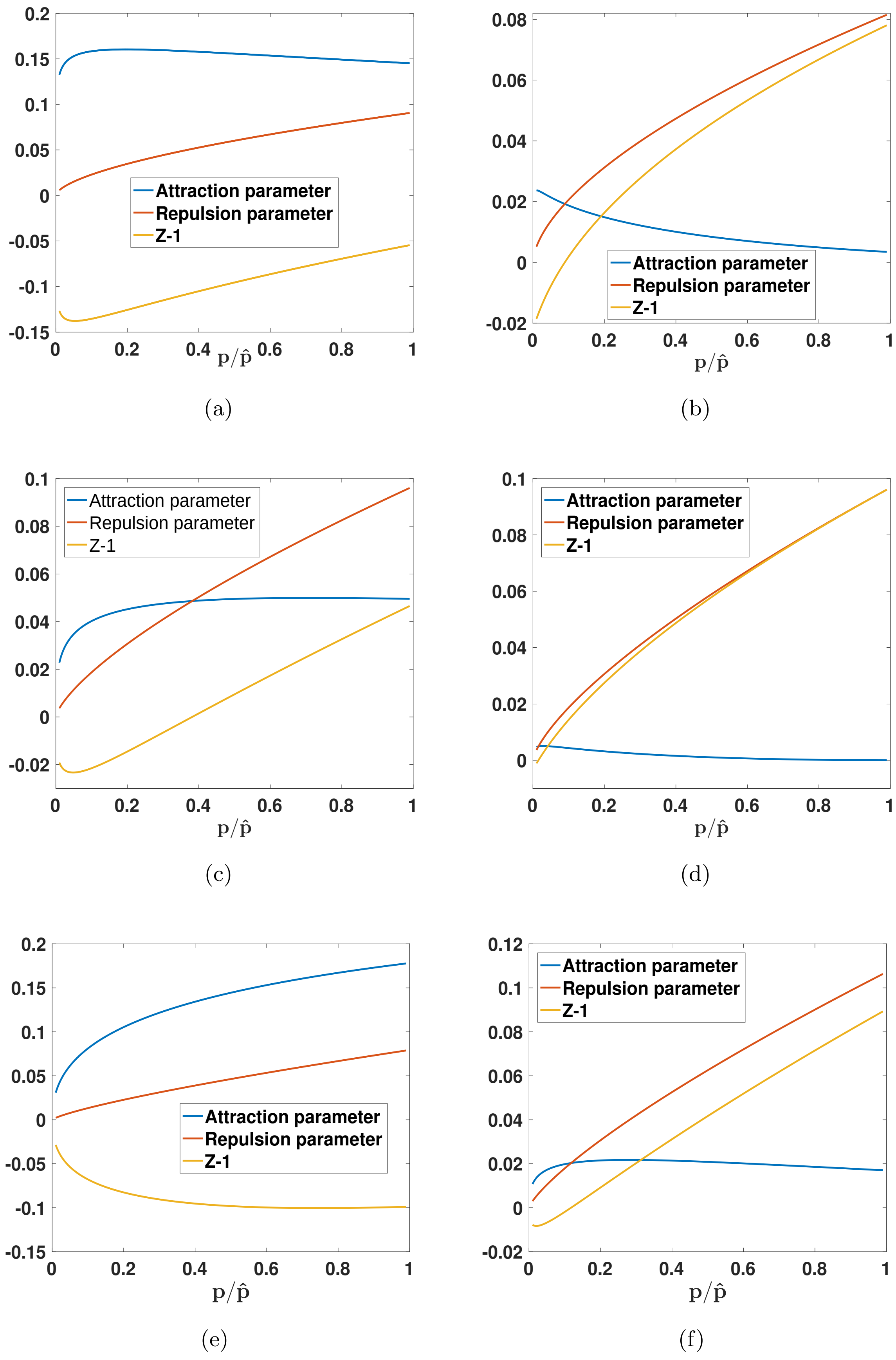

**Fig. 13** **Relative magnitudes of $A$ and $B$ and consequence on whether $Z > 1$ or $Z < 1$ for isentropic expansions and compressions with given stagnation temperature and pressure. (a) Argon, 300 K, 10 MPa; (b) Argon, 1000 K, 30 MPa; (c) Nitrogen, 400 K, 12 MPa; (d) Nitrogen, 1000 K, 30 MPa; (e) Carbon dioxide, 450 K, 10 MPa; (f) Carbon dioxide, 1000 K, 30 MPa.**

## Appendix B: Linearization of the Enthalpy Departure Function

The specific enthalpy $h$ (or enthalpy per mole $\tilde{h} = Wh$) varies from the ideal-gas specific enthalpy $h^*$ (or $\tilde{h}^*$) at the same temperature. Although the present interest is not in two-phase problems, for the SRK case, the gas-phase enthalpy $h_{\rm g}$ and the liquid-phase enthalpy $h_{\rm l}$ each satisfy the following relation:

$$h = \frac{\tilde{h}}{W} = h^*(T) + \frac{1}{W}\big[R_{\rm u}T(Z-1) + \frac{T(da/dT) - a}{b} ln\frac{Z+B}{Z}\big] \tag{B-1}$$

It can be shown from Equation (2.8) that, for a single species,

$$\begin{aligned}
\tilde{a} &\equiv\ 0.42748\frac{(R_{\rm u}T_{\rm c})^2}{p_{\rm c}} \ ; \\
T\frac{da}{dT} &=\ \tilde{a}\left[S^2\frac{T}{T_{\rm c}} - S(S+1)\sqrt{\frac{T}{T_{\rm c}}}\right] \ ; \\
T^2\frac{d^2a}{dT^2} &=\ \frac{\tilde{a}S(S+1)}{2}\sqrt{\frac{T}{T_{\rm c}}}
\end{aligned} \tag{B-2}$$

$$\begin{aligned}
A &=\ \frac{ap}{(R_{\rm u}T)^2} = \frac{\tilde{a}p}{(R_{\rm u}T)^2}\left[(S+1)^2 - 2S(1+S)\sqrt{\frac{T}{T_{\rm c}}} + S^2\frac{T}{T_{\rm c}}\right] \ ; \\
A' &\equiv\ \frac{p}{(R_{\rm u}T)^2}T\frac{da}{dT} = \frac{AT}{a}\frac{da}{dT} = \frac{\tilde{a}p}{(R_{\rm u}T)^2}\left[S^2\frac{T}{T_{\rm c}} - S(S+1)\sqrt{\frac{T}{T_{\rm c}}}\right] \ ; \\
A'' &\equiv\ \frac{p}{(R_{\rm u}T)^2}T^2\frac{d^2a}{dT^2} = \frac{AT^2}{a}\frac{d^2a}{dT^2} = \frac{\tilde{a}p}{(R_{\rm u}T)^2}\frac{S(S+1)}{2}\sqrt{\frac{T}{T_{\rm c}}}
\end{aligned} \tag{B-3}$$

Then,

$$h = h^*(T) + \frac{R_{\rm u}T}{W}\big[Z - 1 + \frac{A' - A}{B} ln\frac{Z+B}{Z}\big] \tag{B-4}$$

For the non-ideal fluid, the volume is not exactly equal to the sum of weighted volumes of the components: $v \neq \Sigma_{\rm j=1}^{\tilde{N}} X_{\rm j}v_{\rm j}$. A similar character occurs for the enthalpy: $\tilde{h} \neq \Sigma_{\rm j=1}^{\tilde{N}} X_{\rm j}\tilde{h}_{\rm j}$.

The enthalpy departure function relation given by Equation (B-4) can be linearized. The result for the enthalpy is

$$\begin{aligned}
h =&\ h^*(T) + \frac{R_{\rm u}T}{W}\big[B - 2A + A'\big] \\
=&\ c_{\rm p}T + \frac{R_{\rm u}\hat{T}}{W}\left[\hat{B} - 2\hat{A}\left[(S+1)^2\frac{\hat{T}}{T} - 2S(1+S)\sqrt{\frac{\hat{T}}{T_{\rm c}}}\sqrt{\frac{\hat{T}}{T}} + S^2\frac{\hat{T}}{T_{\rm c}}\right]\right. \\
&\left. + \hat{A}\left[S^2\frac{\hat{T}}{T_{\rm c}} - S(S+1)\sqrt{\frac{\hat{T}}{T_{\rm c}}}\sqrt{\frac{\hat{T}}{T}}\right]\right]\frac{p}{\hat{p}}
\end{aligned} \tag{B-5}$$

where $\hat{A}$ and $\hat{B}$ are defined using stagnation pressure $\hat{p}$ and stagnation temperature $\hat{T}$.

$$\hat{A} \equiv \frac{\tilde{a}\hat{p}}{(R_{\mathrm{u}}\hat{T})^2} \;\; ; \;\; \hat{B} \equiv \frac{b\hat{p}}{R_{\mathrm{u}}\hat{T}} \tag{B-6}$$

The non-dimensional form is

$$\frac{h}{c_{\mathrm{p}}\hat{T}} = \frac{T}{\hat{T}} + \frac{\gamma - 1}{\gamma}\Bigg[\hat{B} - 2\hat{A}\Bigg[(S+1)^2\frac{\hat{T}}{T} - 2S(1+S)\sqrt{\frac{\hat{T}}{T_{\mathrm{c}}}}\sqrt{\frac{\hat{T}}{T}} + S^2\frac{\hat{T}}{T_{\mathrm{c}}}\Bigg]$$
$$+\hat{A}\Bigg[S^2\frac{\hat{T}}{T_{\mathrm{c}}} - S(S+1)\sqrt{\frac{\hat{T}}{T_{\mathrm{c}}}}\sqrt{\frac{\hat{T}}{T}}\Bigg]\Bigg]\frac{p}{\hat{p}} \tag{B-7}$$

Equation (B-7) can be used to determine temperature. The stagnation enthalpy can be determined given stagnation values for pressure and temperature. It is given by

$$\frac{\hat{h}}{c_{\mathrm{p}}\hat{T}} = 1 + \frac{\gamma - 1}{\gamma}\Bigg[\hat{B} - 2\hat{A}\;(S+1)^2 + 3\hat{A}S(1+S)\sqrt{\frac{\hat{T}}{T_{\mathrm{c}}}} - \hat{A}S^2\frac{\hat{T}}{T_{\mathrm{c}}}\Bigg] \tag{B-8}$$

Equations (B-7) and (B-8) introduce a pressure dependence that does not exist for the ideal gas. Furthermore, these equations indicate that the real-gas enthalpy can exceed the ideal-gas value when $B$ becomes larger than $A$ which occurs as temperature becomes larger. The real-gas enthalpy can fall below the ideal-gas value at more moderate temperatures.

In the next subsection, the wave dynamics for a compressible gas is considered with the purpose of identifying the sound speed which is an important thermodynamic variable in compressible flow.

## Appendix C: Sound Speed

The three variables $p, T$, and $\vec{u}$ can be viewed as governed by the continuity, energy, and momentum equations. Then, coupling with Equations (2.1) and (B-4) also determines $\rho$ and $h$. For the wave dynamics, it is assumed that composition is fixed. Thereby, in the EoS, the quantity $a$ depends only on temperature $T$ and $b$ is fixed. Viscous behavior, body forces, heat conduction, mass diffusion, and turbulent transport are neglected. The following definitions are made: $E$ is the rate of energy addition or conversion per unit mass; at constant composition, consider $p = p(\rho, s)$; $c^2 \equiv \partial p/\partial \rho|_s$ and $e \equiv \partial p/\partial s|_\rho$; $\psi \equiv e/\rho T$ and $\varepsilon \equiv E - \rho T \vec{u} \bullet \nabla s$. Then, the nonlinear wave equation can be developed.

Specifically,

$$\frac{\partial^2 p}{\partial t^2} - c^2\nabla^2 p = \frac{1}{c^2}\frac{\partial c^2}{\partial t}\frac{\partial p}{\partial t} + \psi\frac{\partial \varepsilon}{\partial t} + \varepsilon\frac{\partial \psi}{\partial t} - \frac{1}{c^2}\frac{\partial c^2}{\partial t}\varepsilon\psi + c^2\nabla \bullet (\nabla \bullet (\rho\vec{u}\vec{u})) \quad \text{(C-1)}$$

It is seen from the form of the differential operator in Equation (C-1 ) that the thermodynamic function $c$ is the speed of sound. This conclusion relied only on one thermodynamic condition: a thermodynamic variable is determined, at fixed composition, by the values of two other thermodynamic variables. There has been no assumption about equations of state for density or enthalpy. The velocity $\vec{u}$ can be coupled to pressure $p$ through the Euler momentum equation to close the system for solution.

Now, the speed of sound can be evaluated for our specific equation of state. The differential form of Equation (2.1) is obtained as

$$[3Z^2 - 2Z + A - B - B^2]dZ + [Z - B]dA - [Z + 2BZ + A]dB = 0 \quad \text{(C-2)}$$

Changes in $A$ and $B$ are forced by changes in $T$ and $p$ for constant-composition situations. These in turn cause changes in $Z$. It follows from the EoS that

$$dZ = Z[\frac{dp}{p} + \frac{dv}{v} - \frac{dT}{T}] = Z[\frac{dp}{p} - \frac{d\rho}{\rho} - \frac{dT}{T}] \quad \text{(C-3)}$$

$$dA = A[\frac{dp}{p} - 2\frac{dT}{T}] + A'\frac{dT}{T} \quad \text{(C-4)}$$

$$dB = B[\frac{dp}{p} - \frac{dT}{T}] \quad \text{(C-5)}$$

where $A' \equiv (T/a)(da/dT)A$.

Equations (C-2, C-3 , C-4) and (C-5) may be combined to determine the differential of pressure $dp$ as a function of the temperature and density differentials, $dT$ and $d\rho$. Specifically,

$$\frac{dp}{p} = f(A, B, Z)\frac{d\rho}{\rho} + g(B, Z)\frac{dT}{T} \quad \text{(C-6)}$$

where the definitions are made that

$$f(A, B, Z) \equiv \frac{2Z^3 - Z^2 + AB}{Z^3 - B^2 Z} \quad \text{(C-7)}$$

$$g(B, Z) \equiv \frac{1}{Z - B} - \frac{A'}{Z(Z + B)} \quad \text{(C-8)}$$

From Equation (B-1) the differential relation for enthalpy is derived. Another relation for $dh$ is given by the combined First and Second Law. Matching these two differential

forms yields

$$\frac{dp}{\rho} + Tds = dh = \ c_{\rm p}dT + \frac{R_{\rm u}T}{W}\big[(Z-1)\frac{dT}{T} + dZ - \frac{A-A'}{B}\big(\frac{1}{Z+B} - \frac{1}{Z}\big)dZ$$
$$-\frac{A-A'}{B}\frac{1}{Z+B}dB + \big(\frac{A''}{B}ln\frac{Z+B}{Z}\big)\frac{dT}{T}\big] \tag{C-9}$$

Now, with use of the differential forms given by Equation (C-3, C-4, C-5) for substitution, the following relation is constructed:

$$\frac{dT}{T} = \alpha\frac{dp}{p} + \beta\frac{d\rho}{\rho} + \frac{1}{c_v+\kappa}ds \tag{C-10}$$

where after cancelations $\alpha = 0$ and the following definitions are used:

$$\beta \equiv \ \frac{1}{c_{\rm v}+\kappa}\big[\frac{R_{\rm u}}{W}Z + \frac{a - T\frac{da}{dT}}{bTW}\frac{B}{Z+B}\big] = \frac{(\gamma-1)}{1+\kappa/c_{\rm v}}\big[Z + \frac{A-A'}{Z+B}\big]$$
$$\kappa \equiv \ c_{\rm v}(\gamma-1)\frac{A''}{B}ln\big(\frac{Z+B}{Z}\big) \tag{C-11}$$

Eliminate the temperature differential by substitution from Equation (C-6) with Equation (C-10).

$$\frac{dp}{p} = (f+g\beta)\frac{d\rho}{\rho} + \frac{1}{c_{\rm v}+\kappa}ds \tag{C-12}$$

Thus,

$$c^2 = \ \frac{\partial p}{\partial \rho}\Big|_s = \frac{ZR_{\rm u}T}{W}(f+g\beta) \tag{C-13}$$

For an ideal gas, $Z = 1, A = B = A' = A'' = 0, f = g = 1, \beta = \gamma - 1$ and therefore the well known result, $c^2 = \gamma R_{\rm u}T/W$, follows.

Next, the speed of sound can be evaluated and, when needed, the wave equation (C-1) can be solved. It may be solved together with the Euler form of the momentum equation to determine velocity $\vec{u}$ and pressure $p$. Equation (B-4) governs enthalpy for known temperature or governs temperature for known enthalpy. The density may be determined from Equation (2.1) given $p$ and $T$. These thermodynamic relations are algebraically complicated; when solved with the flow equations, derivatives of these functions are also complicated. In certain regimes of practical relevance, simpler forms can give accurate approximations to these relations. In the following sections, useful, rational simplifying approximations are developed and applied to a few canonical compressible flows. The dependence of the thermodynamic variables on the parameters $A$ and $B$ is linearized.

The nonlinearities in the flow, relating the dependent variables to each other, are fully maintained.

Now, the linearization can be applied to the speed-of-sound function. Within the accuracy of the linear theory, some convenient approximations can be used: $(1+\epsilon_1)/(1+\epsilon_2) \approx (1+\epsilon_1)(1-\epsilon_2) \approx 1+\epsilon_1-\epsilon_2$ where $\epsilon_1$ and $\epsilon_2$ are perturbation quantities with magnitudes smaller than $O(1)$. These standard expansions are used for linearization but the equal sign is used with the approximation understood. Equation (C-13) becomes

$$\frac{\rho c^2}{p} = \frac{\partial(ln\, p)}{\partial(ln\, \rho)}\Big|_s = \gamma + \gamma B + (\gamma-2)A - 2(\gamma-1)A' - (\gamma-1)^2 A'' \qquad \text{(C-14)}$$

Substitutions for $A$ and $B$ in terms of pressure and temperature can be made. Also, an isentropic process is considered so that the derivative in Equation (C-14) becomes the full derivative through the flow field.

$$\begin{aligned}\frac{\rho c^2}{p} = \frac{d(ln\, p)}{d(ln\, \rho)} = \; & \gamma + \gamma\frac{bp}{R_u T} \\ & +(\gamma-2)\frac{\tilde{a}p}{(R_u T)^2}\left[(S+1)^2 - 2S(1+S)\sqrt{\frac{T}{T_c}} + S^2\frac{T}{T_c}\right] \\ & -2(\gamma-1)\frac{\tilde{a}p}{(R_u T)^2}\left[S^2\frac{T}{T_c} - S(S+1)\sqrt{\frac{T}{T_c}}\right] \\ & -(\gamma-1)^2\frac{\tilde{a}p}{(R_u T)^2}\frac{S(S+1)}{2}\sqrt{\frac{T}{T_c}}\end{aligned} \qquad \text{(C-15)}$$

Thus, the linearized form follows:

$$c^2 = \frac{\gamma p}{\rho}[1+\sigma] \qquad \text{(C-16)}$$

where the definition is made that

$$\begin{aligned}\sigma \equiv \; & B + \frac{(\gamma-2)}{\gamma}A - \frac{2(\gamma-1)}{\gamma}A' - \frac{(\gamma-1)^2}{\gamma}A'' \\ = \; & \frac{bp}{R_u T} + \frac{(\gamma-2)}{\gamma}\frac{\tilde{a}p}{(R_u T)^2}\left[(S+1)^2 - 2S(1+S)\sqrt{\frac{T}{T_c}} + S^2\frac{T}{T_c}\right] \\ & -\frac{2(\gamma-1)}{\gamma}\frac{\tilde{a}p}{(R_u T)^2}\left[S^2\frac{T}{T_c} - S(S+1)\sqrt{\frac{T}{T_c}}\right] \\ & -\frac{(\gamma-1)^2}{\gamma}\frac{\tilde{a}p}{(R_u T)^2}\frac{S(S+1)}{2}\sqrt{\frac{T}{T_c}}\end{aligned} \qquad \text{(C-17)}$$

## Appendix D: Relations for Isentropic and Isoenergetic Flows

*Density as a function of pressure:* Using a linear perturbation with Equation (C-15),

it can be approximated that

$$\frac{d(ln\ \rho)}{d(ln\ p)} = \frac{1}{\gamma}\Big[1 - \frac{bp}{R_{\rm u}T} - \frac{\gamma-2}{\gamma}\frac{\tilde{a}p}{(R_{\rm u}T)^2}\big[(S+1)^2 - 2S(1+S)\sqrt{\frac{T}{T_{\rm c}}} + S^2\frac{T}{T_c}\big] + \frac{2(\gamma-1)}{\gamma}\frac{\tilde{a}p}{(R_{\rm u}T)^2}\big[S^2\frac{T}{T_{\rm c}} - S(S+1)\sqrt{\frac{T}{T_{\rm c}}}\big] + \frac{(\gamma-1)^2}{\gamma}\frac{\tilde{a}p}{(R_{\rm u}T)^2}\frac{S(S+1)}{2}\sqrt{\frac{T}{T_{\rm c}}}\Big] \quad \text{(D-1)}$$

The first term on the right (i.e., $1/\gamma$) gives the zeroeth-order term while the other right-side terms provide the first-order correction. Thus, the lowest-order (i.e., zeroeth-order) approximation has $Z = 1$ and $d(ln\ \rho)/d(ln\ p)) = 1/\gamma$. It follows that, to zeroeth order, $\rho/\hat{\rho} = (p/\hat{p})^{1/\gamma}$ and $T/\hat{T} = (p/\hat{p})^{(\gamma-1)/\gamma}$ where $\hat{p}, \hat{\rho}.$ and $\hat{T}$ are stagnation quantities. The zeroeth-order approximation may be substituted in the first-order term with the difference being of second order which has already been declared negligible. Thereby,

$$\frac{d(ln\ \rho)}{d(ln\ p)} = \frac{1}{\gamma}\Big[1 - \frac{b\hat{p}}{R_{\rm u}\hat{T}}\big(\frac{p}{\hat{p}}\big)^{1/\gamma} + \frac{2(\gamma-1)}{\gamma}\frac{\tilde{a}\hat{p}}{(R_{\rm u}\hat{T})^2}\big[S^2\frac{\hat{T}}{T_{\rm c}}\big(\frac{p}{\hat{p}}\big)^{1/\gamma} - S(S+1)\sqrt{\frac{\hat{T}}{T_{\rm c}}}\big(\frac{p}{\hat{p}}\big)^{(3-\gamma)/2\gamma}\big] - \frac{\gamma-2}{\gamma}\frac{\tilde{a}\hat{p}}{(R_{\rm u}\hat{T})^2}\big[(S+1)^2\big(\frac{p}{\hat{p}}\big)^{(2-\gamma)/\gamma} - 2S(1+S)\sqrt{\frac{\hat{T}}{T_{\rm c}}}\big(\frac{p}{\hat{p}}\big)^{(3-\gamma)/2\gamma} + S^2\frac{\hat{T}}{T_{\rm c}}\big(\frac{p}{\hat{p}}\big)^{1/\gamma}\big] + \frac{(\gamma-1)^2}{\gamma}\frac{\tilde{a}\hat{p}}{(R_{\rm u}\hat{T})^2}\frac{S(S+1)}{2}\sqrt{\frac{\hat{T}}{T_{\rm c}}}\big(\frac{p}{\hat{p}}\big)^{(3-\gamma)/2\gamma}\Big] \quad \text{(D-2)}$$

Separation of variables and integration yields

$$\frac{\rho}{\hat{\rho}} = C\big(\frac{p}{\hat{p}}\big)^{1/\gamma} exp\big[-\lambda_{\rm b}\big(\frac{p}{\hat{p}}\big)^{1/\gamma} + \lambda_1\big(\frac{p}{\hat{p}}\big)^{1/\gamma} - \lambda_2\big(\frac{p}{\hat{p}}\big)^{(2-\gamma)/\gamma} + \lambda_3\big(\frac{p}{\hat{p}}\big)^{(3-\gamma)/2\gamma}\big] \quad \text{(D-3)}$$

where $C$ is the constant of integration and

$$\lambda_{\rm b} \equiv \hat{B}\ ; \qquad \lambda_1 \equiv S^2\hat{A}\frac{\hat{T}}{T_{\rm c}}\ ;$$

$$\lambda_2 \equiv -\frac{1}{\gamma}(S+1)^2\hat{A}\ ; \qquad \lambda_3 \equiv \frac{\gamma+1}{\gamma}S(S+1)\hat{A}\sqrt{\frac{\hat{T}}{T_{\rm c}}} \quad \text{(D-4)}$$

$$\hat{A} \equiv \frac{\tilde{a}\hat{p}}{(R_{\rm u}\hat{T})^2}\ ; \qquad \hat{B} \equiv \frac{b\hat{p}}{R_{\rm u}\hat{T}} \quad \text{(D-5)}$$

Now, the exponential term is expanded to the needed order to obtain

$$\frac{\rho}{\hat{\rho}} = C\big(\frac{p}{\hat{p}}\big)^{1/\gamma}\big[1 - \lambda_{\rm b}\big(\frac{p}{\hat{p}}\big)^{1/\gamma} + \lambda_1\big(\frac{p}{\hat{p}}\big)^{1/\gamma} - \lambda_2\big(\frac{p}{\hat{p}}\big)^{(2-\gamma)/\gamma} + \lambda_3\big(\frac{p}{\hat{p}}\big)^{(3-\gamma)/2\gamma}\big] \quad \text{(D-6)}$$

$C$ is determined by setting $\rho = \hat{\rho}$ when $p = \hat{p}$. Upon expansion to a linear form, $C = 1+\lambda_{\rm b}-\lambda_1+\lambda_2-\lambda_3$. Substitution for $C$ into equation (D-6) followed by multiplication

yields

$$\frac{\rho}{\hat{\rho}} = \big(\frac{p}{\hat{p}}\big)^{1/\gamma} + (\lambda_1 - \lambda_{\rm b})\big[\big(\frac{p}{\hat{p}}\big)^{2/\gamma} - \big(\frac{p}{\hat{p}}\big)^{1/\gamma}\big]$$
$$-\lambda_2\big[\big(\frac{p}{\hat{p}}\big)^{(3-\gamma)/\gamma} - \big(\frac{p}{\hat{p}}\big)^{1/\gamma}\big] + \lambda_3\big[\big(\frac{p}{\hat{p}}\big)^{(5-\gamma)/2\gamma} - \big(\frac{p}{\hat{p}}\big)^{1/\gamma}\big]$$
$$= \big(\frac{p}{\hat{p}}\big)^{1/\gamma}\Big[1 + \Lambda_1\big(\frac{p}{\hat{p}}\big)\Big] \tag{D-7}$$

where $\Lambda_1(p/\hat{p})$ is defined as follows:

$$\Lambda_1 \equiv (\lambda_1 - \lambda_{\rm b})\big[\big(\frac{p}{\hat{p}}\big)^{1/\gamma} - 1\big] - \lambda_2\big[\big(\frac{p}{\hat{p}}\big)^{(2-\gamma)/\gamma} - 1\big] + \lambda_3\big[\big(\frac{p}{\hat{p}}\big)^{(3-\gamma)/2\gamma} - 1\big] \tag{D-8}$$

$\Lambda_1$ is the fractional variation of the real-gas isentropic relation between density and pressure from the ideal-gas isentropic relation between those same variables. In the case where the stagnation pressure and temperature are given and fixed in the comparison, a factor $\hat{Z}$ is still needed to account for the difference in stagnation density.

*Enthalpy as a function of pressure:* The enthalpy can be obtained by a simple integration for an isentropic process: $h = \int dh = \int (1/\rho) dp$. First, a relation is obtained for the reciprocal of density as a function of pressure.

$$\frac{\hat{\rho}}{\rho} = \big(\frac{p}{\hat{p}}\big)^{-1/\gamma} + (\lambda_{\rm b} - \lambda_1)\big[1 - \big(\frac{p}{\hat{p}}\big)^{-1/\gamma}\big]$$
$$+\lambda_2\big[\big(\frac{p}{\hat{p}}\big)^{(1-\gamma)/\gamma} - \big(\frac{p}{\hat{p}}\big)^{-1/\gamma}\big] - \lambda_3\big[\big(\frac{p}{\hat{p}}\big)^{(1-\gamma)/2\gamma} - \big(\frac{p}{\hat{p}}\big)^{-1/\gamma}\big]$$
$$= \big(\frac{p}{\hat{p}}\big)^{-1/\gamma}[1 - \Lambda_1\big(\frac{p}{\hat{p}}\big)] \tag{D-9}$$

Now, integration with use of stagnation values to set the constant yields

$$\hat{h} - h = \frac{u^2}{2} = \frac{\gamma}{\gamma - 1}\big(\frac{\hat{p}}{\hat{\rho}}\big)\Big[1 - \big(\frac{p}{\hat{p}}\big)^{(\gamma-1)/\gamma} + \Lambda_2\big(\frac{p}{\hat{p}}\big)\Big] \tag{D-10}$$

where the function $\Lambda_2(p/\hat{p})$ is defined as follows to encapsulate first-order terms.

$$\Lambda_2 \equiv (\lambda_1 - \lambda_{\rm b})\big[\frac{\gamma - 1}{\gamma}\frac{p}{\hat{p}} - \big(\frac{p}{\hat{p}}\big)^{(\gamma-1)/\gamma} + \frac{1}{\gamma}\big] - \lambda_2\big[(\gamma - 1)\big(\frac{p}{\hat{p}}\big)^{1/\gamma} - \big(\frac{p}{\hat{p}}\big)^{(\gamma-1)/\gamma} + 2 - \gamma\big]$$
$$+\lambda_3\big[\frac{2(\gamma - 1)}{\gamma + 1}\big(\frac{p}{\hat{p}}\big)^{(\gamma+1)/2\gamma} - \big(\frac{p}{\hat{p}}\big)^{(\gamma-1)/\gamma} + \frac{3 - \gamma}{\gamma + 1}\big] \tag{D-11}$$

$\Lambda_2$ is the fractional variation of the real-gas isentropic relation between kinetic energy per unit mass $(u^2/2)$ and pressure from the ideal-gas isentropic relation between those same variables.

Equation (B-8) for stagnation enthalpy $\hat{h}$ can be used to substitute into Equation (D-10) to determine enthalpy $h$. The left-side of Equation (B-8) is the ratio of real-gas stagnation enthalpy to ideal-gas stagnation enthalpy. Thus, an increase in $A$ gives a relative increase to the ideal-gas value while an increase in $B$ gives a relative increase to the real-gas value. In an ideal gas, the only energy is the kinetic (translational and rotational) energy of molecules while the real gas also has an energy associated with intermolecular forces.

*Temperature as a function of pressure:* The combination of Equation (D-10) yielding the enthalpy and Equation (B-7) yielding the linearized enthalpy departure function allows the determination of temperature as a function of pressure. Specifically,

$$
\begin{aligned}
\frac{T}{\hat{T}} &= 1 + \frac{\gamma-1}{\gamma}\Bigg[\hat{B}(1-\frac{p}{\hat{p}}) + 2\hat{A}(S+1)^2((\frac{p}{\hat{p}})^{1/\gamma}-1) \\
&\quad +3\hat{A}S(1+S)\sqrt{\frac{\hat{T}}{T_{\mathrm{c}}}}(1-(\frac{p}{\hat{p}})^{(\gamma+1)/2\gamma}) + \hat{A}S^2\frac{\hat{T}}{T_{\mathrm{c}}}(\frac{p}{\hat{p}}-1)\Bigg] \\
&\quad -\hat{Z}\Bigg[1-(\frac{p}{\hat{p}})^{(\gamma-1)/\gamma} + (\lambda_1-\lambda_{\mathrm{b}})[\frac{\gamma-1}{\gamma}\frac{p}{\hat{p}} - (\frac{p}{\hat{p}})^{(\gamma-1)/\gamma} + \frac{1}{\gamma}] \\
&\quad -\lambda_2[(\gamma-1)(\frac{p}{\hat{p}})^{1/\gamma} - (\frac{p}{\hat{p}})^{(\gamma-1)/\gamma} + 2-\gamma] \\
&\quad +\lambda_3[\frac{2(\gamma-1)}{\gamma+1}(\frac{p}{\hat{p}})^{(\gamma+1)/2\gamma} - (\frac{p}{\hat{p}})^{(\gamma-1)/\gamma} + \frac{3-\gamma}{\gamma+1}]\Bigg] \\
&= 1 + \frac{\gamma-1}{\gamma}\Bigg[\hat{B}(1-\frac{p}{\hat{p}}) + 2\hat{A}(S+1)^2((\frac{p}{\hat{p}})^{1/\gamma}-1) \\
&\quad +3\hat{A}S(1+S)\sqrt{\frac{\hat{T}}{T_{\mathrm{c}}}}(1-(\frac{p}{\hat{p}})^{(\gamma+1)/2\gamma}) \\
&\quad +\hat{A}S^2\frac{\hat{T}}{T_{\mathrm{c}}}(\frac{p}{\hat{p}}-1)\Bigg] - \hat{Z}\Big[1-(\frac{p}{\hat{p}})^{(\gamma-1)/\gamma} + \Lambda_2(\frac{p}{\hat{p}})\Big] \qquad \text{(D-12)}
\end{aligned}
$$

*Velocity as a function of pressure:* For isoenergetic flow, $\hat{h} - h = u^2/2$ where $u$ is the velocity. From Equation (D-10), it follows that

$$
\begin{aligned}
\frac{u^2}{2c_{\mathrm{p}}\hat{T}} = \frac{\hat{h}-h}{c_{\mathrm{p}}\hat{T}} &= \hat{Z}\Big[1-(\frac{p}{\hat{p}})^{(\gamma-1)/\gamma} + (\lambda_1-\lambda_{\mathrm{b}})[\frac{\gamma-1}{\gamma}\frac{p}{\hat{p}} - (\frac{p}{\hat{p}})^{(\gamma-1)/\gamma} + \frac{1}{\gamma}] \\
&\quad -\lambda_2[(\gamma-1)(\frac{p}{\hat{p}})^{1/\gamma} - (\frac{p}{\hat{p}})^{(\gamma-1)/\gamma} + 2-\gamma] \\
&\quad +\lambda_3[\frac{2(\gamma-1)}{\gamma+1}(\frac{p}{\hat{p}})^{(\gamma+1)/2\gamma} - (\frac{p}{\hat{p}})^{(\gamma-1)/\gamma} + \frac{3-\gamma}{\gamma+1}]\Big] \\
&= \hat{Z}\Big[1-(\frac{p}{\hat{p}})^{(\gamma-1)/\gamma} + \Lambda_2(\frac{p}{\hat{p}})\Big] \qquad \text{(D-13)}
\end{aligned}
$$

Thereby,

$$\frac{u}{(2c_\mathrm{p}\hat{T})^{1/2}} = \hat{Z}^{1/2}\Big[1-\big(\frac{p}{\hat{p}}\big)^{(\gamma-1)/\gamma}+(\lambda_1-\lambda_\mathrm{b})\big[\frac{\gamma-1}{\gamma}\frac{p}{\hat{p}}-\big(\frac{p}{\hat{p}}\big)^{(\gamma-1)/\gamma}+\frac{1}{\gamma}\big]$$
$$-\lambda_2\big[(\gamma-1)\big(\frac{p}{\hat{p}}\big)^{1/\gamma}-\big(\frac{p}{\hat{p}}\big)^{(\gamma-1)/\gamma}+2-\gamma\big]$$
$$+\lambda_3\big[\frac{2(\gamma-1)}{\gamma+1}\big(\frac{p}{\hat{p}}\big)^{(\gamma+1)/2\gamma}-\big(\frac{p}{\hat{p}}\big)^{(\gamma-1)/\gamma}+\frac{3-\gamma}{\gamma+1}\big]\Big]^{1/2}$$
$$= \hat{Z}^{1/2}\Big[1-\big(\frac{p}{\hat{p}}\big)^{(\gamma-1)/\gamma}+\Lambda_2\big(\frac{p}{\hat{p}}\big)\Big]^{1/2} \quad \text{(D-14)}$$

*Sound speed as a function of pressure:* From Equation (C-16),

$$c^2 = \frac{\gamma p}{\rho}[1+\sigma] = \frac{\gamma p}{\rho}\big[1+B+\frac{\gamma-2}{\gamma}A-\frac{2(\gamma-1))}{\gamma}A'-\frac{(\gamma-1)^2}{\gamma}A''\big]$$
$$= \frac{\gamma p}{\rho}\Big[1+(\lambda_\mathrm{b}-\lambda_1)\big(\frac{p}{\hat{p}}\big)^{1/\gamma}+\frac{\lambda_2(2-\gamma)}{\gamma}\big(\frac{p}{\hat{p}}\big)^{(2-\gamma)/\gamma}-\frac{(3-\gamma)\lambda_3}{2}\big(\frac{p}{\hat{p}}\big)^{(3-\gamma)/2\gamma}\Big]$$
$$= \frac{\gamma\hat{p}}{\hat{\rho}}\big(\frac{p}{\hat{p}}\big)^{(\gamma-1)/\gamma}\Bigg[1-\lambda_\mathrm{b}+\lambda_1-\lambda_2+\lambda_3+2(\lambda_\mathrm{b}-\lambda_1)\big(\frac{p}{\hat{p}}\big)^{1/\gamma}$$
$$+\frac{2\lambda_2}{\gamma}\big(\frac{p}{\hat{p}}\big)^{(2-\gamma)/\gamma}-\frac{(5-\gamma)\lambda_3}{2}\big(\frac{p}{\hat{p}}\big)^{(3-\gamma)/2\gamma}\Bigg]$$
$$= \frac{\gamma\hat{p}}{\hat{\rho}}\big(\frac{p}{\hat{p}}\big)^{(\gamma-1)/\gamma}\Big[1+\Lambda_3\big(\frac{p}{\hat{p}}\big)\Big] \quad \text{(D-15)}$$

where the function $\Lambda_3(p/\hat{p})$ is defined as follows to encapsulate the first-order terms.

$$\Lambda_3 \equiv -\ \lambda_\mathrm{b}+\lambda_1-\lambda_2+\lambda_3+2(\lambda_\mathrm{b}-\lambda_1)\big(\frac{p}{\hat{p}}\big)^{1/\gamma}$$
$$+\frac{2\lambda_2}{\gamma}\big(\frac{p}{\hat{p}}\big)^{(2-\gamma)/\gamma}-\frac{(5-\gamma)\lambda_3}{2}\big(\frac{p}{\hat{p}}\big)^{(3-\gamma)/2\gamma} \quad \text{(D-16)}$$

$\Lambda_3$ is the fractional variation of the real-gas isentropic relation between sound speed squared and pressure from the ideal-gas isentropic relation between those same variables. In the case where the stagnation pressure and temperature are given and fixed in the comparison, a factor $\hat{Z}$ is still needed to account for the difference in stagnation density value from the ideal-gas value.

Upon normalization, it may be written that

$$\frac{c^2}{2c_\mathrm{p}\hat{T}} = \frac{\gamma-1}{2}\hat{Z}\big(\frac{p}{\hat{p}}\big)^{(\gamma-1)/\gamma}\Bigg[1-\lambda_\mathrm{b}+\lambda_1-\lambda_2+\lambda_3$$
$$+2(\lambda_\mathrm{b}-\lambda_1)\big(\frac{p}{\hat{p}}\big)^{1/\gamma}+\frac{2\lambda_2}{\gamma}\big(\frac{p}{\hat{p}}\big)^{(2-\gamma)/\gamma}-\frac{(5-\gamma)\lambda_3}{2}\big(\frac{p}{\hat{p}}\big)^{(3-\gamma)/2\gamma}\Bigg]$$
$$= \frac{\gamma-1}{2}\hat{Z}\big(\frac{p}{\hat{p}}\big)^{(\gamma-1)/\gamma}\Big[1+\Lambda_3\big(\frac{p}{\hat{p}}\big)\Big] \quad \text{(D-17)}$$

and

$$\frac{c}{(2c_{\rm p}\hat{T})^{1/2}} = \left(\frac{\gamma-1}{2}\right)^{1/2}\hat{Z}^{1/2}\big(\frac{p}{\hat{p}}\big)^{(\gamma-1)/2\gamma}\Big[1 + A_3\big(\frac{p}{\hat{p}}\big)\Big]^{1/2} \tag{D-18}$$